\documentclass[a4paper,12pt]{article}
\usepackage[
bookmarks=true,bookmarksnumbered=true,%
linkcolor=blue,anchorcolor=blue,urlcolor=blue,
]{hyperref}
\usepackage{amsfonts,amsthm,amsmath,amssymb}
\usepackage{graphicx}
\usepackage{braket}
\usepackage{bm}
\usepackage{float}
\usepackage{mathrsfs}
\usepackage{booktabs}
\usepackage{makeidx}
\usepackage{color}
\usepackage{tikz}
\numberwithin{equation}{section}
\newcommand{\be}{\begin{equation}}
\newcommand{\ee}{\end{equation}}
\newcommand{\bea}{\begin{eqnarray}}
\newcommand{\eea}{\end{eqnarray}}

\newcommand{\Tr}{\text{Tr}}

\newcommand{\arctanh}{\text{arctanh}}

\newcommand{\ba}{\begin{eqnarray}}
\newcommand{\ea}{\end{eqnarray}}

\newcommand{\mcl}{\mathcal}

\newcommand{\f}{\frac}
\newcommand{\s}{\sqrt}

\newcommand{\ep}{\epsilon}

 \def\f {\frac}

 \def \ep{\epsilon}

\begin{document}

\begin{titlepage}
\thispagestyle{empty}

\begin{flushright}

OU-HET 986

\end{flushright}


\begin{center}
\noindent{{\textbf{Holographic Complexity for disentangled states}}}\\
\vspace{2cm}
Tokiro Numasawa $^{1,2}$\vspace{1cm}


\vskip 1em
{\it 
$^1$
Department of Physics, McGill University,\\ 
3600 rue University, Montr\'eal, Qu\'ebec, Canada H3A 2T8}

{\it
$^2$ 
Department of Physics, Graduate School of Science,\\
Osaka university, Toyonaka 560-0043, Japan
}

\vskip 2em
\end{center}

\begin{abstract}
In this paper we consider the maximal volume and the action, which are conjectured to be gravity duals of the complexity, in the black hole geometries with end of the world branes.
These geometries are duals of boundary states in CFTs which have small real space entanglement.
When we raise the black hole temperature while keeping the cutoff radius, black hole horizons or end of the world branes come in contact with the cutoff surface.
In this limit, holographic entanglement entropy reduces to $0$.
We studied the behavior of the volume and the action.
We found that the volume reduces to $0$ in this limit.
The behavior of the action depends on their regularization.
We study the implication of these results to the reference state of the holographic complexity both in the complexity = volume or the complexity = action conjectures.
\end{abstract}

\end{titlepage}
\newpage
\tableofcontents

\section{Introduction}
Recently, it is found that there are deep connections between quantum information theory and quantum gravity through holography\cite{tHooft:1993dmi,Susskind:1994vu}, especially in the AdS/CFT correspondence\cite{Maldacena:1997re}.
The holographic entanglement entropy formula \cite{Ryu:2006bv,Hubeny:2007xt} relates entanglement entropy in conformal field theories(CFTs) to bulk minimal (extremal) codimension $2$ surfaces in gravity.
These relations leads to the understanding that the bulk geometry appears from the boundary entanglement\cite{Swingle:2009bg,VanRaamsdonk:2010pw}.

We can probe the black hole interiors by entanglement entropy but that saturates at finite time in finite entropy systems\cite{Hartman:2013qma}. 
On the other hand, black hole interiors expand beyond this saturation time.
The codimension $1$ maximal volume that ends on the asymptotic boundary can detect the growth of interiors after the saturation of entanglement\cite{Susskind:2014rva}.
The volume is now conjectured to be dual of complexity in dual conformal field theories\cite{Susskind:2014rva,Stanford:2014jda}.
Through the hard wall approximation of interface solutions, it is also argued that the maximal volume is the gravity dual of quantum information metric \cite{MIyaji:2015mia}.
The on shell action evaluated on so called Wheeler de Witt (WdW) patch also shows the same behavior at late time, and they are also conjectured to be a dual of complexity in CFTs\cite{Brown:2015bva,Brown:2015lvg}.
Based on tensor network consideration, it is also argued that the Liouville action corresponds to the path integral complexity\cite{Miyaji:2016mxg,Caputa:2017urj,Caputa:2017yrh,Czech:2017ryf}, which represents redundancy to prepare states through the Euclidean path integral.
Recently, the connection of the gravity action and quantum circuits based on path integrals are proposed\cite{Takayanagi:2018pml}. 

The complexity of states is defined by the number of elementary gates applied to a reference state\cite{Susskind:2014moa}.
This means that we need to specify the "reference state" and "gate set" to define the complexity, and complexity depends on the choice of them.
Holographic formula, on the other hand, gives a way to evaluate complexity using geometric  quantities in AdS.
This means that the holographic formula chooses one particular reference state and a gate set.
Therefore, in principle, we can specify the reference state and gate set from dual "definition" of complexity from gravity side.
Here we focus on the reference state.
The natural candidate of reference state is a product state, which have no real space entanglement\cite{Susskind:2014rva}.
Therefore, to test the reference state from gravity calculation, the key point is how to realize such a small entanglement state in holographic setup.
In CFTs, there are states with small real space entanglement.
They are called as boundary states\cite{CARDY1989581,Cardy:2004hm}.
This is a state that represents the existence of boundary which keeps the half of conformal symmetry.
These states have small entanglement and used to cut UV entanglement \cite{Ohmori:2014eia,Cardy:2016fqc} near the entangling surface, to express a projection operation in CFTs\cite{Rajabpour:2015uqa,Numasawa:2016emc} or to describe approximately the ground states of gapped Hamiltonians\cite{Calabrese:2005in,Cardy:2017ufe}.
Recently is is also argued that boundary CFTs are used to construct an analog of qubits in CFTs\cite{VanRaamsdonk:2018zws}.
These boundary conditions have rich structures and especially the dual geometries highly depend on the choice of boundary conditions.
There is a holographic model that is called as AdS/BCFT correspondence proposed by Takayanagi\cite{Takayanagi:2011zk} and explored further in \cite{Fujita:2011fp}.
In this proposal, we assume that the effects of the boundaries in CFT side are expressed as end of the world  (ETW) branes in gravity side.
In this paper, we consider both of the complexity=volume and the complexity = action conjecture in AdS$_3$/BCFT$_2$ context where BCFTs have spacelike boundaries.
The volume and the action with in AdS/BCFT are also studied in \cite{Flory:2017ftd} where BCFTs have timelike boundaries.

An important point is that product states are actually not in a Hilbert space of any relativistic QFT\cite{Witten:2018zxz}.
Reflecting this fact, the norms of boundary states are infinite.
In gravity side, the change of UV entanglement breaks the asymptotic AdS boundary condition.
This is related to the fact that the identity operator is not in a trace class in QFTs.
The identity operator is the infinite temperature limit of the Gibbs ensembles and it is singular both in field theory and in gravity.
In this paper, we consider the AdS/CFT correspondence with cutoff\cite{Susskind:1998dq}.
By UV/IR relation, this cutoff is a suitable UV cutoff in CFT and an IR cut off in gravity. 
The area of the cutoff surface in gravity is interpreted as the quantum information which we have in CFT with the suitable UV regularization.
The prescription to calculate holographic entanglement entropy ending on a generic bulk surface are considered in \cite{Miyaji:2015yva}.
We consider the limit that the horizon or the ETW brane meet with the cutoff surface.
Such a limit is sometimes considered\cite{Miyaji:2014mca,Bredberg:2010ky,Nomura:2017fyh,Reynolds:2016rvl}.
This limit is considered in \cite{Miyaji:2014mca} for Euclidean black holes to argue that boundary states correspond to trivial spacetimes.
In this limit, there are no spacetime within the cutoff surface and holographic entanglement entropy reduces to $0$.
In this paper, we consider this limit in Lorentzian signature and consider the volume and the WdW patch action in this limit.

The result is as follows:

\noindent
1. In the complexity=volume case, if there are no entanglement, the states have $0$ complexity. This suggests that the reference state of complexity for CV case is product states.

\noindent
2. In the complexity=action case, the result depends on how to regularize the Wheeler de Witt patch action.
We consider two regularizations that are considered in the literature.
When we choose the Wheeler de Witt patch that ends on the AdS boundary and then introduce a cutoff, we obtain a UV divergence for product state limits.
On the other hand, when we consider the Wheeler de Witt patch that ends on the cutoff surface, then complexity reduces to $0$ for product state limits.
These are not affected by  the choice of the affine parametrization ambiguity of joint terms \cite{Lehner:2016vdi,Carmi:2016wjl} or length scale of counter terms\cite{Lehner:2016vdi,Chapman:2018dem,Chapman:2018lsv}.

The organization of this paper is as follows.
In section 2, we review the property of correlation function and entanglement in boundary states.
Then, we review the basic aspects of AdS/BCFT and see the structure of spacetime.
In section 3, we study the volume and the action on the eternal black holes.
We also study study them in the AdS/BCFT setup with tensionless ETW branes, which are obtained by the orbifold of the eternal black holes.
We study the behavior of the volume and the action in the limit that the horizon radius becomes the same with the cutoff radius where holographic entanglement entropy reduces to $0$.
In section 4, we study the volume and the action in the AdS/BCFT setup with nonzero tension ETW branes.
For positive tension case, we consider the limit that the radius of black hole horizon becomes the same with the cutoff radius where only the geometry behind the horizon remains. 
For the geometry with negative tension ETW branes, we consider the limit where the ETW brane contacts with the cutoff surface at $t=0$ where holographic entanglement entropy reduces to $0$.

\section{Entanglement structure of boundary states}
\subsection{single sided CFT case}
In this section we study the real space correlation in boundary states $\ket{B}$ in conformal field theories.
More details are discussed in \cite{Miyaji:2014mca}\cite{Guo:2017rji}.
Because boundary states have infinite norms, we consider the regularized version of boundary states $\ket{\psi} = e^{- \delta \cdot H} \ket{B}$ for a smearing parameter $\delta$ and take it to be cutoff scale $\epsilon$ of the CFT.
 
Let us consider a two point function of two local operators:
\be
\braket{ O(x_1) O(x_2)} = \bra{\psi} O(x_1) O(x_2) \ket{\psi}.
\ee
This correlation function is evaluated on the infinite strip $[-\delta,\delta]\times \mathbb{R}$ where the Euclidean time $\tau$ runs from $-\delta$ to $\delta$ and we use $z$ as a coordinate for this strip.  
Operators are located on the reflection symmetric line $\tau = 0$.
Using conformal mapping $w = e^{\frac{\pi}{2\delta}z}$, the strip geometry is mapped to the upper half plane (UHP).
The cross ratio $x$ on the UHP becomes 
\be
x = \frac{(w_1-\bar{w}_1)(w_2-\bar{w}_2)}{(w_1-w_2)(\bar{w}_1-\bar{w}_2)} =- \frac{1}{\sinh^2\f{\pi}{4\delta} (x_1-x_2)}.
\ee
Therefore, when $|x_1-x_2| \gg \delta$, the cross ratio becomes $x \approx 0$.
This means that two point functions factorize to the products of one point functions:
\be
\braket{O(x_1)O(x_2)} \sim \braket{O(x_1)}\braket{O(x_2)}, \label{eq:bdyev}
\ee
and they do not depend on the separation of them.
This implies that there are no real space correlation in boundary states.

We can also estimate the real space correlation by entanglement entropy. 
We consider the entanglement of an infinite interval.
Then, this reduces to the one point function of the twist operator:
\be
\bra{\psi} \sigma_n(0) \ket{\psi} = \tilde{c}_n\Big( \f{4 \delta}{\pi \ep} \Big) ^{-\Delta_n}.
\ee
Therefore, entanglement entropy becomes
\be
S_A = \f{c}{6} \log \f{4\delta}{\pi \ep} + \log g + \f{1}{2} c_1' \label{eq:lineeebdy}.
\ee
Here $c_1'$ is a non universal constant which does not depend on the choice of boundary states and related to $\tilde{c}_1'$ by $\tilde{c}_1'= c_1'/2 +  \log g$\cite{Calabrese:2009qy}.
This $\log g$ is so called boundary entropy and $g$ is the disk amplitude $g = \braket{0|B}$\cite{PhysRevLett.67.161}.
Note that this expression can be trusted only when $\delta \gg \ep$.
Therefore, we can not trust the expression (\ref{eq:lineeebdy}).
Nevertheless, we can estimate the nonexistence of logarithmic divergent term in (\ref{eq:lineeebdy}) when we take $\delta \sim \ep$.

Let us consider qubit (or spin) systems with $\mcl{H}_{tot} = \mcl{H}^{ \otimes n}$ where $\mcl{H}$ is spanned by $\ket{0}$ and $\ket{1}$. 
An example of product states $\ket{\psi_0} \in \mcl{H}_{tot}$ is given by
\be
\ket{\psi_0} = \ket{0}\ket{0} \cdots \ket{0}.
\ee
The important feature of this state is that two point functions are factorized to the products of one point functions because there is no entanglement:
\be
\bra{\psi_0} O(i)O(j) \ket{\psi_0} = \bra{\psi_0} O(i) \ket{\psi_0}\bra{\psi_0} O(j) \ket{\psi_0},
\ee
where the lavel $i,j$ is a lavel of cites on which qubits are located.
This is a same property which is satisfied in boundary state in (\ref{eq:bdyev}).
Therefore, we can see a boundary state as a CFT representation of a product state.

\subsection{two sided CFT case}
In two sided systems with maximal horizontal entanglement, the analog of product states is one with no vertical entanglement\cite{Susskind:2014rva}.
In qubit systems, this is a product of $n$ EPR pairs each of which is shared between the two sides:
\be
\ket{\psi} = \{1/\s{2}(\ket{0}_L\ket{0}_R + \ket{1}_L \ket{1}_R)\}^{\otimes n}
\ee 
We see that the properties of Bell pairs are satisfied by the following thermofield double state:
\be
\ket{EPR} = \f{1}{\s{Z(4\delta)}} \sum_n e^{-2 \delta\cdot E_n}\ket{E_n}_L \otimes \ket{E_n}_R^{CPT},\label{eq:TFD}
\ee
where we take $\delta $ to be a cutoff scale $\ep$.

Fist, the two point functions of two left (right) operators in EPR pairs factorizes to a product of one point functions:
\be
\bra{\psi} O_L(i)O_L(j) \ket{\psi} - \Tr_L( \rho_L O_L(i))\Tr_L( \rho_L O_L(j)) =  0.
\ee
where $\rho_L = \frac{1}{2}(\ket{0}_L\bra{0}_L + \ket{1}_L\bra{1}_L)$.
In CFTs, we obtain
\ba
\bra{EPR} O(x_1)O(x_2) \ket{EPR} = \Big(\f{\pi}{ 4\delta}\Big)^{4 \Delta} \f{1}{|\sinh\f{\pi}{4\delta}(x_1-x_2)|^{4\Delta}} \notag \\
\sim  \Big(\f{\pi}{ 2\delta}\Big)^{4 \Delta} e^{-\f{\pi \Delta}{\delta}|x_1-x_2|},
\ea
which means that the two point function factorizes to the product of one point functions\footnote{In conformal field theories on $S^1 \times \mathbb{R}$, one point functions vanish.}.

Next, if we consider the two point functions of left and right operators, that have non zero correlation only when they are located on the same point:
\be
\bra{\psi} O_L(i)O_R(j) \ket{\psi}- \Tr_L( \rho_L O_L(i))\Tr_R( \rho_R O_L(j)) = 0 \qquad (i\neq j).
\ee
The two sided correlator in CFTs is
\ba
\bra{EPR} O(i2\delta+x_1)O(x_2)\ket{EPR} &=& \Big(\f{\pi}{4 \delta}\Big)^{4 \Delta} \f{1}{|\cosh\f{\pi}{4\delta}(x_1-x_2)|^{4\Delta}}
 \notag \\
&\sim& 
\begin{cases}
\Big(\f{\pi}{ 2\delta}\Big)^{4 \Delta} e^{-\f{\pi \Delta}{\delta}|x_1-x_2|} &  |x_1 - x_2| \gg \delta \\
\Big(\f{\pi}{ 4\delta}\Big)^{4 \Delta}  & x_1 \sim x_2
\end{cases}.
\ea
Again we can confirm that the behavior is resemble to that of EPR pairs.
Entanglement Renyi entropy is given by
\be
\bra{EPR} \sigma_n(i 2\delta + x_1) \sigma_{-n}(x_1) \ket{EPR} = c_n \Big(\frac{\pi \ep}{4\delta} \Big) ^{4\Delta_n}.
\ee
and entanglement entropy becomes
\be
S_A = \frac{c}{3} \log \frac{4\delta}{\pi \ep} + c_1 '.
\ee
This is essentially the twice of the boundary state entanglement entropy (\ref{eq:lineeebdy}).
When we take $\delta$ to be the cutoff scale $\epsilon$, the logarithmic divergent disappears.
This suggests that the thermo field double states (\ref{eq:TFD}) share the same properties with the EPR pairs in qubits systems.




\subsection{Holographic dual of EPR pairs}
We start from the dual of maximally entangled states without vertical entanglement, which are simpler than boundary states.
The dual of $\ket{EPR} = \f{1}{\s{Z(4\delta)}} \sum_n e^{-2 \delta\cdot E_n}\ket{E_n}_L \otimes \ket{E_n}_R^{CPT}$ is given by the eternal AdS black holes\cite{Maldacena:2001kr}.
The gravity dual of local EPR pairs are also given in \cite{Numasawa:2016emc}.
The exterior coordinate of the BTZ black hole is given by
\be
ds^2 = - \f{r^2 - r_H^2}{l_{AdS}^2}dt^2 + \f{l_{AdS}^2}{r^2 - r_H^2}dr^2 + r^2 d\phi^2,
\ee
where $l_{AdS}$ is the AdS radius and $r_H$ is the horizon radius. $\phi$ is related to the boundary space coordinate $x$ via $\phi = x /l_{AdS}$.
The black hole temperature is given by $\beta = 2 \pi l_{AdS}^2 /r_H$.

The relation with CFT cutoff is determined as follows.
First, when we take $r \gg r_H$, the metric takes the following form approximately:
\be
ds ^2 \sim - \f{r^2}{l_{AdS}^2} dt^2 + \f{l_{AdS}^2}{r^2} dr^2 + r^2 d\phi^2 .
\ee
Then, when we take $z = l_{AdS}^2/r$, we obtain the usual Poincare patch metric:
\be
ds^2 \sim l_{AdS}^2 \f{dz ^2 -dt^2+ dx^2  }{z^2}.
\ee
Therefore, the cutoff scale is given by $z = z_0= \epsilon$.
In the original coordinate, the cutoff is $r_0 = l_{AdS}^2/\epsilon$.
The above is applicable only for $r_0 \gg r_H$.
When we take the temperature to be a cutoff scale, we assume that we can trust this expression $r_0 = l_{AdS}^2/\epsilon$ and  take the limit $r_H \to 2\pi / \epsilon$ that corresponds to $r_H \to r_0$
\footnote{Strictly speaking, disentangled state is not in the Hilbert space in relativistic field theories  because all states in relativistic field theories have the same UV divergence in entanglement entropy\cite{Witten:2018zxz}.
In gravity side, all dual geometries have the same asymptotic AdS region that causes the universal UV divergence in holographic entanglement entropy.
Therefore the horizon does not meet with the asymptotic AdS boundary no matter how high we take temperature to be.
Therefore we need some regularization in both side.
We assume the existence of holographic regularization and we identify putting cutoff surface in AdS with taking the local EPR pair limit in the regularized Hilbert space.}
.
In other word, we identify putting entanglement entropy in the thermo field double states ( or boundary states) to $0$ in CFT and putting cutoff on horizon in gravity:
\be
S_A = \frac{c}{3} \log \frac{4\delta}{\pi\ep} + c_1 ' \to 0 \qquad \text{in CFT} \qquad \leftrightarrow \qquad r_H \to r_0 \qquad \text{in gravity} \label{eq:identify}
\ee
Actually, we can confirm that there are no holographic entanglement entropy in this limit.
Entanglement entropy for EPR pairs(thermofield double states) of the half line $A: x>0$ for both copies of the CFT can be calculated by holographically\cite{Ryu:2006bv},
\be
S_A = \f{l_{AdS}}{4G_N}\int_{r_H}^{r_0} \f{dr}{\s{r^2-r_0^2}}  = \f{l_{AdS}}{2G_N} \log \f{r_0+\s{r_0^2 - r_H^2}}{r_H},
\ee  
which becomes $0$ in $r_H \to r_0$ limit\footnote{This calculation only mean that there are no entanglement which is leading in $1/N$ expansion.}.
This suggest that we can identify both limit in (\ref{eq:identify}).
We can say that the trivial space is associated to the EPR pairs without vertical entanglement at $t=0$, which is argued in \cite{Miyaji:2014mca}.
After the time evolution, entanglement are generated and they create the interior of the black holes.

\subsection{Holographic dual of Boundary states}
In this subsection we consider the holographic dual of (regularized) boundary states $e^{-\delta\cdot H}\ket{B}$.
We assume the AdS/BCFT setup \cite{Takayanagi:2011zk}\cite{Fujita:2011fp}.
Dual geometry is given by black holes with end-of-the-world (ETW) brane \cite{Hartman:2013qma}\cite{Almheiri:2018ijj}.
After reviewing the basics  AdS/BCFT prescription, we explain the geometry with cutoff scale temperature.

The action for holographic model of BCFT is given as follows:
\be
S = \f{1}{16 \pi G_N} \int \s{-g }(R - 2\Lambda) + \f{1}{8\pi G_N} \int_{brane} \s{-\gamma} (K -  T) .
\ee
We impose the Neumann condition on the brane. The equation of motion on the end-of-the world (ETW) brane is given by
\be
K_{ab} - h_{ab}K = - T h_{ab}.
\ee
We can think of this equation as the junction condition \cite{Israel:1966rt} with nothing.
By taking the trace, we obtain
\be
K = 2T.
\ee
The trajectory of the ETW brane in left outside region is given by
\be
r(t) = \f{r_H}{\s{1 - (Tl_{AdS})^2 }} \s{1 - (T l_{AdS})^2 \tanh^2\f{r_H t}{l_{AdS}^2}}
\ee
The induced metric on the ETW brane is 
\be
ds_{brane}^2 = - \f{r_H^4}{l_{AdS}^2}  \Big(\f{Tl_{AdS}}{1 - (Tl_{AdS})^2}\Big) ^2  \f{1}{r(t)^2}\f{1}{\cosh^4 \f{r_H t}{l_{AdS}^2}}  dt ^2 + r(t)^2 d\phi^2.
\ee
\begin{figure}[ht]
\begin{minipage}{0.32\hsize}
\begin{center}
\includegraphics[width=4.5cm]{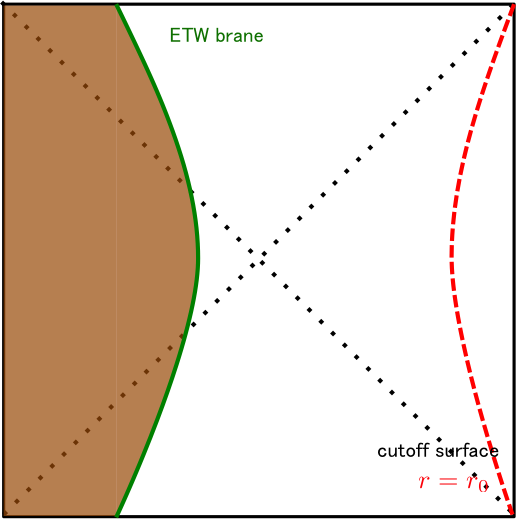}
\end{center}
\end{minipage}
\begin{minipage}{0.32\hsize}
\begin{center}
\includegraphics[width=4.5cm]{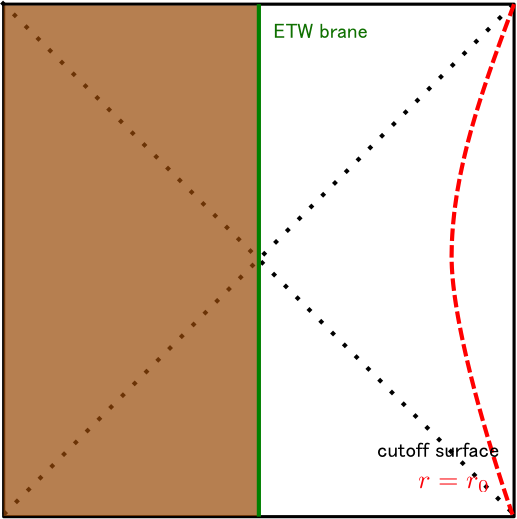}
\end{center}
\end{minipage}
\begin{minipage}{0.32\hsize}
\begin{center}
\includegraphics[width=4.5cm]{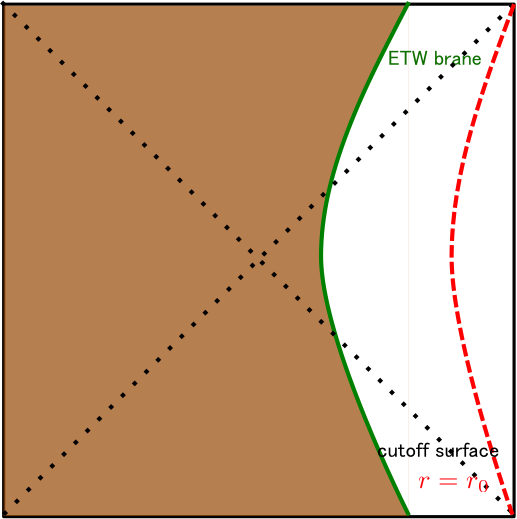}
\end{center}
\end{minipage}
\caption{The left picture is the configuration of the ETW brane with positive tension.
The middle picture is the configuration of the tensionless ETW brane, which is obtained as the orbifold of eternal black holes.
The right picture is the configuration of the ETW brane with negative tension }
\label{fig:ETWconfig}
\end{figure}
The configurations of the ETW branes are depicted in Figure \ref{fig:ETWconfig}.

Here we argue that in this case the relation $\frac{c}{3}\log \frac{4\delta}{\pi \ep} \to 0 \leftrightarrow r_H \to r_0$ also holds.
First, we calculate holographic entanglement entropy \cite{Ryu:2006bv,Hubeny:2007xt} for a half interval in BTZ black strings with ETW branes.
Holographic entanglement entropy at $t=0$ is calculated by the length of the geodesics on $t=0$ slice that starts from the cutoff surface and end on the ETW brane\cite{Takayanagi:2011zk,Fujita:2011fp}\footnote{In this case, we can consider geodesics that end on the ETW brane. This corresponds to the fact that one point function can have expectation value in BCFT.}. The answer is
\ba
S_A &=&\int_{r_0}^{r_H} dr  \f{l_{AdS}}{\s{r^2 - r_H^2}} +   \int_{r_H}^{r_B} dr  \f{l_{AdS}}{\s{r^2 - r_H^2}} \notag \\
&=&\f{l_{AdS}}{4G_N} \log \frac{r_0 + \sqrt{r_0^2 - r_H^2}}{r_H} + \f{l_{AdS}}{4G_N} \log \f{1+Tl_{AdS}}{\s{1-(Tl_{AdS})^2}}  \notag \\
&=& \f{l_{AdS}}{4G_N} \log \frac{r_0 + \sqrt{r_0^2 - r_H^2}}{r_H}+\f{l_{AdS}}{4G_N} \arctanh(Tl_{AdS}) \label{eq:bdyentropy},
\ea
where $r_B = r(0) = \f{r_H}{\s{1-(Tl_{AdS})^2}}$.
The second contribution in this expression of holographic entanglement entropy is exactly equal to the boundary entropy $\log g$ obtained from the holographic disk partition function in AdS/BCFT setup \cite{Takayanagi:2011zk}\cite{Fujita:2011fp} when we take the limit $r_H \to r_0$.
Therefore, this is equivalent to putting $\frac{c}{3}\log \frac{4\delta}{\pi \ep} + c_1' = 0$ in (\ref{eq:lineeebdy})
Let us keep $r_H$ to be different from $r_0$ but to be the same order i.e. to keep the ratio $r_H/r_0$ to be finite.
Then (\ref{eq:bdyentropy}) does not have UV divergence but becomes finite.
This corresponds to the fact that the $\log \f{\delta}{\epsilon}$ term in (\ref{eq:lineeebdy}) is finite when we take $\delta$ to be a cutoff scale $\epsilon$.
This is not true for the volume or the action case as we will see in later chapters.

Now, boundary states with UV reguralization have finite entanglement entropy after removing the logarithmic term $\f{c}{3} \log \f{4\delta}{\pi \ep}$ and non universal constant term $c_1'$. 
This finite entanglement builds new region that is a part of the right side of the black hole according to the  relation of entanglement and bulk geometry \cite{VanRaamsdonk:2010pw}.
It is interesting that the entanglement in boundary state create the region that are not causally connected to the cutoff surface, which will be related to entanglement wedge reconstruction \cite{Headrick:2014cta}.

Holographic entanglement entropy for an interval with length $x$ is also easily computed.
There are two configurations for geodesics.
The first configuration is two disconnected geodesics that end on the ETW brane.
For this case, holographic entanglement entropy is given by 
\be
S_{dis} = \f{l_{AdS}}{2G_N} \log \frac{r_0 + \sqrt{r_0^2 - r_H^2}}{r_H}+\f{l_{AdS}}{2G_N} \arctanh(Tl_{AdS}) 
\ee
The second configuration is the connected geodesics.
For this case, holographic entanglement entropy is given by 
\be
S_{con} = \f{l_{AdS}}{4G_N} \text{arccosh}\Big(\frac{r_0^2}{r_H^2} \cosh \frac{r_Hx}{l_{AdS}^2} + 1\Big)
\ee
The value of holographic entanglement entropy is given by the minimal one of 
\be
S_A = \min\{ S_{dis},S_{con}\}
\ee
In $r_H \to r_0$ limit, we obtain 
\ba
S_{dis} &=& \frac{l_{AdS}}{2G_N}\arctanh(Tl_{AdS}) \notag \\
S_{con} &=& \frac{l_{AdS}}{4G_N}\frac{r_0x}{l_{AdS}^2}
\ea

When we take $T=0$, this geometry is the one that is obtained in \cite{Hartman:2013qma}\footnote{$T=0$ cases have stringy realization.
For example, in the duality between type IIA sting  and ABJM theory\cite{Aharony:2008ug}, O8 plane with $8$ D8 branes realizes such a ETW brane\cite{Fujita:2011fp}.
After lifting to M theory that becomes Horava-Witten wall \cite{Horava:1996ma}.
}.
In this case ETW brane is obtained as the fixed point of orbifold of the original eternal black hole.
There are no entanglement in the leading of $1/N$ expansion at $t=0$ when we take $r_H \to r_0$ limit.
Therefore, we can think of them as a holographic realization of product states.

When $T < 0$, the configuration of ETW brane is again given by 
\be
r(t) = \f{r_H}{\s{1 - (Tl_{AdS})^2 }} \s{1 - (T l_{AdS})^2 \tanh^2\f{r_H t}{l_{AdS}^2}}, \label{eq:collapseshell}
\ee
but now the ETW brane is located on the left side in which they are causally connected to the asymptotic boundary.
In this $T<0$ case, the boundary entropy becomes negative.
From the view of entanglement/geometry connection, negative boundary entropy eliminate a part of geometry from the original left asymptotic region and also from the inside of black holes.
This configuration can be seen as a model of gravitational collapse.
Initially the shell represented by the ETW brane is located on the $r = \frac{r_H}{\s{1 - (Tl_{AdS})^2 }}$ surface. 
The radius of the ETW brane shrinks along the trajectory (\ref{eq:collapseshell}) and finally makes a black hole.

In negative tension case we can not pull the horizon to the cutoff surface.
Because the maximal radius of the ETW brane on the right outside region is $r = \frac{r_H}{\s{1 - (Tl_{AdS})^2 }}$,
the maximal horizon radius is $r_H = \s{1 - (Tl_{AdS})^2} r_0 $, which is smaller than the cutoff radius $r_0$.
In this limit $r_H \to \s{1 - (Tl_{AdS})^2} r_0$ the ETW brane comes in contact with the cutoff surface at $t=0$.
There are no holographic entanglement and product states are realized in the dual CFT with a suitable UV cutoff.


\section{Volume and Action for duals of local EPR pairs}
Before going to study the complexity for holographic duals of boundary states, we study holographic complexities for two sided black holes, which are dual to maximal entanglement without vertical entanglement.
The results are applicable to single sided case with tensionless ETW brane because geometric quantities are essentially given by the half of those of eternal black holes.
We concentrate on the calculation at $t=0$.

\subsection{Volume cases}
The complexity = volume conjecture suggests the following correspondence:
\be
C_V = \f{V}{G_N l_{AdS}}
\ee
where $V$ is the maximal volume that end on the time slice of AdS boundary where CFT states are defined.

A useful coordinate for the calculation of volumes is the following one:
\be
ds^2 = - \f{r^2 - r_H^2}{l_{AdS}^2}dt^2 + \f{l_{AdS}^2}{r^2 - r_H^2}dr^2 + r^2 d\phi^2,
\ee
where $l_{AdS}$ is the AdS radius and $r_H$ is the horizon radius. $\phi$ is related to the boundary space coordinate $x$ via $\phi = x /l_{AdS}$.
We use $L$ to represent the period of $x$ direction $\int dx$, which is the $1$ dimensional volume of space direction that the boundary theory lives in.
The black hole temperature is given by $\delta = 2 \pi l_{AdS}^2 /r_H$, which is identified with the smearing parameter of the thermo field double state(\ref{eq:TFD}) in CFT side.

It is easy to evaluate the volume at $t = 0$ in this coordinate.
The answer is 
\ba
V = l_{AdS}\int d\phi d r \s{\f{r^2}{r^2-r_H^2}} &=& 2 l_{AdS} \cdot \f{L}{l_{AdS}} \int _{r_H}^{r_0} dr \f{r}{\s{r^2 - r_H^2}} \notag \\
&=& 2 L \s{r_0^2 -r_H^2}, \label{volumeEPR}
\ea
and the holographic complexity defined by the volume is 
\ba
C_V = \f{V}{G_N l_{AdS}} &=& \f{2L}{G_N l_{AdS}} \s{r_0^2 - r_H^2} \notag \\
&=& 3 c \f{L}{\epsilon} \s{1- \Big(\f{\pi \ep}{2\delta} \Big)^2} \label{eq:EBHvolume}
.
\ea
where we used Brown-Henneaux central charge $c = \f{3 l_{AdS}}{2 G_N}$\cite{Brown:1986nw}.
We can see that $C_V$ vanishes when we take $\delta = \pi\epsilon/2$ i.e. $r_H \to r_0$ limit.

When we consider single sided case, the volume is given by the half of (\ref{volumeEPR}), that becomes again $0$ in $r_H \to r_0$ limit.
This shows that the holographic complexity vanishes for a state without entanglement.

In the volume case, when we take $r_H$ to be a cutoff scale $r_0$ but to be different from $r_0$, the volume (\ref{eq:EBHvolume}) is UV divergent.
This is different from entanglement cases (\ref{eq:lineeebdy}) and (\ref{eq:bdyentropy}), where they are finite in this parameter regime.

\subsection{Action cases}
The complexity$=$ action conjecture proposes the following correspondence\cite{Brown:2015bva}\cite{Brown:2015lvg}:
\be
C = \frac{\mathcal{A}_{WdW}}{\pi \hbar}
\ee
where $\mathcal{A}_{WdW}$ is the value of on shell action $S_{WdW}$ evaluated on so called Wheeler de Witt (WdW) patch.
Here, the action functional on WdW patch is defined by
\be
S_{WdW} = \f{1}{16\pi G_N} \int _{WdW} \s{-g} (R -2 \Lambda) + \f{1}{8 \pi G_N} \int_{\text{bdy}} \s{-\gamma} K + S_{\text{null bdy}} + S_{joint} + S_{ct}
\ee
where first two terms are usual Einstein-Hilbert action and Gibbons-Hawking term \cite{Gibbons:1976ue} for boundaries which are spacelike or timelike.
On the other hand, $S_{\text{null bdy}}$ is the action for null boundaries and given by
\be
S_{null bdy} = \f{1}{8 \pi G_N}\int \kappa \s{\gamma} d\lambda d\theta^{d-2}.
\ee
where $\kappa$ satisfies $k^{\beta}\nabla_{\beta} k^{\alpha} = \kappa k^{\alpha}$ for a null generator $k^{\alpha}$ on the null surfaces, $\lambda$ is a parameter on the null generator $k^{\alpha}$, and $\theta ^A (A= 1 ,\cdots, d-2)$ is constant on each null generator. 
This term can be set to $0$ by taking the affine parametrization on the null boundary.
In this paper we use affine parametrization on the boundary and set this term to be $0$.
$S_{joint}$ are terms that comes from the joints of boundaries\cite{Lehner:2016vdi} and take the form of 
\be
S_{joint} = \f{1}{8\pi G_N} \int \s{\gamma} d\theta^{d-2} a , 
\ee
where $a$ is  given by the logarithm of the inner products of normal vectors of jointing surfaces when one of them is null.
This term is ambiguous under the rescaling of affine parametrization.
There is also a counter term\cite{Lehner:2016vdi}\cite{Chapman:2018dem}\cite{Chapman:2018lsv} on the null boundary 
\be
S_{ct} = \f{1}{8 \pi G_N} \int \lambda d\theta^{d-2} \s{\gamma} \Theta \log (l_{ct} \Theta) ,
\ee
where $\Theta = \partial_{\lambda}\log \s{\gamma}$. If we include this counter terms, they eliminate the affine parametrization dependence of $S_{joint}$.


To evaluate the action, the following Kruscal metric is useful:
\be
ds^2 = -l_{AdS}^2 \f{ 4du dv}{(1+uv)^2} + r_H^2\f{(1-uv)^2}{(1+uv)^2} d\phi^2
\ee
In this coordinate, $uv = -1$ corresponds to the boundary and $uv = 1$ is the singularity of BTZ black hole.
On the other hand, $uv=0$ is the black hole horizon and $u = v = 0$ is the bifurcation surface.
The right out side region is given by $v>0, u<0$. 

The on-shell value of Einstein-Hilbert term  is evaluated as 
\ba
\f{1}{16 \pi G_N}\int \s{-g} (R - 2 \Lambda) &=& \f{1}{16 \pi G_N}\int \s{-g} (R - 2\f{d-2}{2d} R) \notag \\
&=& \f{1}{16 \pi G_N}\f{2}{d} \int \s{-g}R \notag \\
&=& -\f{1}{8 \pi G_N} \f{d-1}{l_{AdS}^2} \int \s{-g}.
\ea
Therefore it reduces to the spacetime volume of the WdW patch.
Especially, when we consider BTZ black holes $(d=3)$, this becomes
\be
\f{1}{16 \pi G_N}\int \s{-g} (R - 2 \Lambda) = -\f{1}{4 \pi G_Nl_{AdS}^2}  \int \s{-g}
\ee
Note that the bulk contribution is always negative, because the space time volume is always positive.

\begin{figure}[ht]
\begin{minipage}{0.5\hsize}
\begin{center}
\includegraphics[width=5cm]{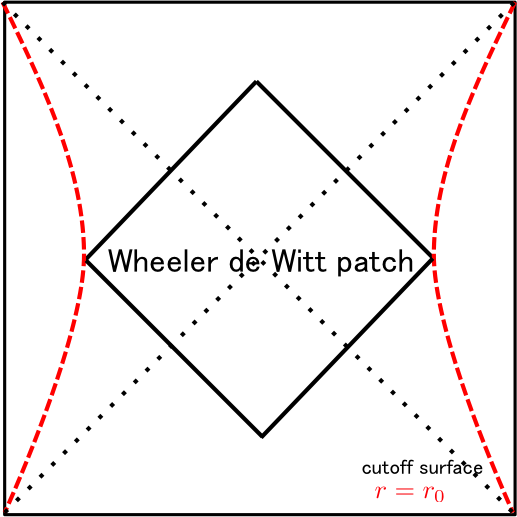}
\end{center}
\end{minipage}
\begin{minipage}{0.5\hsize}
\begin{center}
\includegraphics[width=5cm]{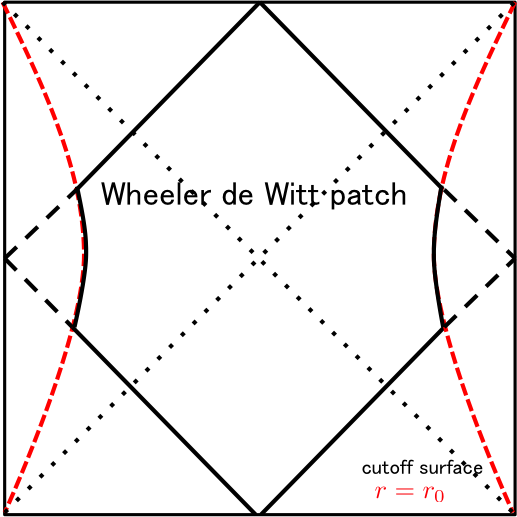}
\label{fig:choice2}
\end{center}
\end{minipage}
\caption{Two regularizations of the WdW patch. The left picture is the first regularization, in which the WdW patch ends on the cutoff surface.
The right picture is the second regularization, in which the WdW patch end on the asymptotic AdS boundary.}
\label{fig:choices}
\end{figure}

It should be noted that the regularization of the action on Wheeler de Witt patch is not unique\cite{Carmi:2016wjl}.
In this paper we consider two type of regularization that are considered in \cite{Carmi:2016wjl}.
The first one is the WdW patch that ends on the cutoff surface.
The second regularization is the WdW patch that ends on the asymptotic AdS boundary and then introduce a cutoff surface.
They are shown in fig. \ref{fig:choices}.

\subsubsection{Regularization 1}
In this case, there are no boundary which is timelike or spacelike.
Therefore, we only need to calculate the Einstein-Hilbert term, the joint term and the counter term action.
The WdW patch with cutoff surface is given by the following region:
\be
-\s{\f{r_0-r_H}{r_0+r_H}} \le u,v \le \s{\f{r_0-r_H}{r_0+r_H}}.
\ee
The Einstein-Hilbert term contribution is 
\ba
S_{bulk} = - \f{L}{2\pi G_N l _{AdS}} \Big( r_0 - \f{r_H^2}{r_0} \Big)
\ea
The joint term contribution is 
\ba
S_{joint} = -\f{L}{4\pi G_N l_{AdS}} r_0 \log \f{\alpha \tilde{\alpha} l_{AdS}^2}{r_0^2 - r_H^2} + \f{L}{4\pi G_N l_{AdS}} \f{r_H^2}{r_0} \log  \Big( \f{r_0^2}{r_H^2}\f{\alpha \tilde{\alpha} l_{AdS}^2}{r_0^2 - r_H^2} \Big).
\ea
where 
$\alpha, \tilde{ \alpha}$ is the normalization of affine null normals on each null surface.
The counter term contribution is 
\ba
S_{ct} = \f{L}{4 \pi G_N l_{AdS}} \Big( 2 r_0 + r_0 \log \f{\alpha \tilde{\alpha} l_{ct}^2}{r_0^2}\Big) - \f{L}{4 \pi G_N l_{AdS}} \Big( 2\f{r_H^2}{r_0^2} + \f{r_H^2}{r_0} \log \f{\alpha \tilde{\alpha} l_{ct}^2}{(r_H^2/r_0)^2}\Big),
\ea
where $l_{ct}$ is a dimensionful parameter that is needed to introduce the counter term\cite{Lehner:2016vdi,Chapman:2018dem,Chapman:2018lsv}.
Therefore, holographic complexity for the action is

\ba
C_A &=& \f{S_{bulk} + S_{joint} + S_{counter}}{\pi} \notag \\
&=& \f{L}{4\pi^2 G_N l _{AdS}} \Big[- 2r_0 + 2\f{r_H^2}{r_0} - r_0 \log \f{\alpha \tilde{\alpha} l_{AdS}^2}{r_0^2 - r_H^2} + \f{r_H^2}{r_0} \log  \Big( \f{r_0^2}{r_H^2}\f{\alpha \tilde{\alpha} l_{AdS}^2}{r_0^2 - r_H^2}   \Big)\Big)  \notag \\
 &&+\Big( 2 r_0 + r_0 \log \f{\alpha \tilde{\alpha} l_{ct}^2}{r_0^2}\Big) - \Big( 2\f{r_H^2}{r_0^2} + \f{r_H^2}{r_0} \log \f{\alpha \tilde{\alpha} l_{ct}^2}{(r_H^2/r_0)^2}\Big) \Big] \notag \\ 
 &=& \f{L}{4\pi^2 G_N l _{AdS}} \Big( r_0 \log \f{(r_0^2 - r_H^2) l_{ct}^2}{r_0^2 l_{AdS}^2} - \f{r_H^2}{r_0}\log  \f{(r_0^2 - r_H^2) l_{ct}^2}{r_H^2 l_{AdS}^2} \Big) \label{eq:EternalChoice1}
\ea
This becomes $0$ when we take $r_H \to r_0$ limit.
Note that this results does not depend on the choice of $l_{ct}$.

We can also consider the action without the counter term, which now depends on the choice of the affine parametrization on the null boundaries:
\ba
C_A &=& \f{S_{bulk} + S_{joint} }{\pi} \notag \\
&=& \f{L}{4\pi^2 G_N l _{AdS}} \Big(- 2r_0 + 2\f{r_H^2}{r_0} - r_0 \log \f{\alpha \tilde{\alpha} l_{AdS}^2}{r_0^2 - r_H^2} + \f{r_H^2}{r_0}  \log  \Big( \f{r_0^2}{r_H^2}\f{\alpha \tilde{\alpha} l_{AdS}^2}{r_0^2 - r_H^2}   \Big)\Big)   \notag 
\ea
This also becomes $0$ when we take $r_H \to r_0$ limit.
This is because the contribution from $S_{bulk}, S_{joint}$ and $S_{ct}$ independently reduce to $0$.
Therefore, even when we omit the counter term, this result does not depend on the choice of the affine parametrization on the null boundaries.
These facts suggest that the reference state of complexity is chosen to a product state in this regularization of WdW patch.


\subsubsection{Regularization 2}
Next we consider the second regularization, in which the WdW patch end on the asymptotic boundary and then cutoff surfaces are introduced on $r = r_0$.
The WdW patch are given by the region $-1 \le u,v \le 1$ which also satisfy $r \le r_0$.
In this regularization, the cutoff surface is a timelike boudnary.
Therefore, we also need to compute the contribution from the Gibbons Hawking term.
The Einstein Hilbert term contribution is 
\be
S_{EH} = -\f{L}{2 \pi G_N l_{AdS}} r_H \Big( 1 + \f{2r_H \s{r_0-r_H}}{\s{r_0+r_H} + \s{r_0-r_H}} - \f{\s{r_0^2 -r_H^2}}{(\s{r_0+r_H} - \s{r_0-r_H})^2} \log \f{r_0-r_H}{r_0+r_H} \Big).
\ee
The Gibbons-Hawking term contribution is
\be
S_{GH} = \f{L}{4\pi G_N l_{AdS}} \f{(2r_0^2 - r_H^2)}{r_H} \log \f{r_0 + r_H}{r_0 - r_H}. 
\ee
The contribution from joint term is
\be
S_{joint} = -\f{L }{4 \pi G_Nl_{AdS}} r_0 \log \f{\alpha \tilde{\alpha} l_{AdS}^2}{r_0^2-r_H^2}.
\ee
The the counter term contribution is
\be
S_{ct} = \f{L}{4 \pi G_N l_{AdS}} \Big (2 r_0 + r_0 \log\f{\alpha\tilde{\alpha} l_{ct}^2}{r_0^2} \Big).
\ee
Therefore, complexity becomes 
\ba
C_A &=& \f{S_{EH} + S_{GH} + S_{joint} + S_{ct}}{\pi} \notag \\
&=& \f{L}{4 \pi^2 G_N l_{AdS}}  \Big( -2r_H - \f{4r_H \s{r_0-r_H}}{\s{r_0+r_H} + \s{r_0-r_H}} + \f{2r_H\s{r_0^2 -r_H^2}}{(\s{r_0+r_H} - \s{r_0-r_H})^2} \log \f{r_0-r_H}{r_0+r_H}  \notag \\ 
&&+ \f{(2r_0^2 - r_H^2)}{r_H} \log \f{r_0 + r_H}{r_0 - r_H} -r_0 \log \f{\alpha \tilde{\alpha} l_{AdS}^2}{r_0^2-r_H^2} + 2 r_0 + r_0 \log\f{\alpha\tilde{\alpha} l_{ct}^2}{r_0^2} \Big) \notag \\
&=&  \f{L}{4 \pi^2 G_N l_{AdS}}  \Big(2 (r_0-r_H) - \f{4r_H \s{r_0-r_H}}{\s{r_0+r_H} + \s{r_0-r_H}} + \f{2r_H\s{r_0^2 -r_H^2}}{(\s{r_0+r_H} - \s{r_0-r_H})^2} \log \f{r_0-r_H}{r_0+r_H}  \notag \\ 
&&+ \f{(2r_0^2 - r_H^2)}{r_H} \log \f{r_0 + r_H}{r_0 - r_H}   + r_0 \log\f{ (r_0^2 -r _H^2)l_{ct}^2}{r_0^2 l_{AdS}^2} \Big)
\ea
In $r_H \to r_0$ limit, the complexity takes a simple form:
\ba
C_A &=& \frac{L}{2\pi^2 G_N l_{AdS}} \Big[ r_0 \log \f{2l_{ct}}{l_{AdS}}   \Big] \notag \\
&=& \f{c}{3\pi^2}  \f{L}{\epsilon} \log \f{2l_{ct}}{l_{AdS}} 
\ea
If we do not include the counter term, we obtain
\ba
C_A  &=& \frac{L}{2\pi^2 G_N l_{AdS}} r_0\log\f{2 r_0}{\s{\alpha \tilde{\alpha}}l_{AdS}}  -\f{L}{2 \pi G_N l_{AdS}} r_0 \notag \\
 &=& \frac{c}{3\pi^2} \f{L}{\epsilon} \log \f{2 l_{AdS}}{\s{\alpha \tilde{\alpha}} \epsilon} - \f{c}{3\pi^2 }\f{L}{\epsilon} 
\ea
which is $\f{1}{\epsilon} \log\f{1}{\epsilon} $ divergent behavior that found earlier in the regularization without counter terms \cite{Carmi:2016wjl}.
In this regularization of WdW patch complexity have UV divergence for (EPR) pairs with vertical entanglement.

\subsection{Volume and action for $T=0$ boundary states}
The solution with ETW branes for $T=0$ (tensionless) is obtained by the $\mathbb{Z}_2$ orbifold of eternal black holes.
Correspondingly, the volume and action is obtained by the half of the results for eternal black holes.
Here we summarize the results for tensionless boundary states case.
The volume becomes the half of (\ref{eq:EBHvolume}):
\ba
C_V =  \f{L}{G_N l_{AdS}} \s{r_0^2 - r_H^2}. \label{eq:t0v}
\ea
The action with regularization $1$, in which the WdW patch ends on the cutoff surface, is the half of (\ref{eq:EternalChoice1}):
\ba
C_A =  \f{L}{8\pi^2 G_N l _{AdS}} \Big( r_0 \log \f{(r_0^2 - r_H^2) l_{ct}^2}{r_0^2 l_{AdS}^2} - \f{r_H^2}{r_0}\log  \f{(r_0^2 - r_H^2) l_{ct}^2}{r_H^2 l_{AdS}^2} \Big).\label{eq:t0a1}
\ea
The action with regularization $2$, in which the WdW patch ends on the cutoff surface, is the half of 

\ba
C_A &=&\f{L}{8\pi^2 G_N l_{AdS}}  \Big(2 (r_0-r_H) - \f{4r_H \s{r_0-r_H}}{\s{r_0+r_H} + \s{r_0-r_H}} + \f{2r_H\s{r_0^2 -r_H^2}}{(\s{r_0+r_H} - \s{r_0-r_H})^2} \log \f{r_0-r_H}{r_0+r_H}  \notag \\ 
&&+ \f{(2r_0^2 - r_H^2)}{r_H} \log \f{r_0 + r_H}{r_0 - r_H}   + r_0 \log\f{ (r_0^2 -r _H^2)l_{ct}^2}{r_0^2 l_{AdS}^2} \Big) \label{eq:t0a2}
\ea
$r_H \to r_0$ limit behaviors are also the same.
In this limit, the volume and the action with regularization $1$ reduces to $0$.
On the other hand, the action with regularization $2$ becomes
\be
C_A = \frac{L}{4\pi^2 G_N l_{AdS}} \Big[ r_0 \log \f{2l_{ct}}{l_{AdS}}   \Big] . \label{eq:t0a2limit}
\ee
We will confirm this in the next section by taking the tensionless limit of general results for non zero tension cases.









\section{Volume and Action for Boundary state with non-zero boundary entropy}

In this section we consider holographic dual of boundary state with non-zero boundary entropy.
As is the case with the thermofield double state, we have the Hawking-Page transition\cite{Hawking:1982dh} in AdS/BCFT setup\cite{Takayanagi:2011zk}\cite{Fujita:2011fp} when we vary the black hole radius.
Because we are interested in the high temperature limit, we only consider the black hole phases.
\subsection{Volume cases}
In this subsection, we consider the CV conjecture.
The volume  is given by that of  the $t=0$ codimension $1$ slice that start from the cutoff surface and end on the ETW brane.
As is the case with holographic entanglement entropy, we can separate the volume to two part.
The first one is the volume of the $t=0$ slice that starts from the cutoff surface and end on the bifurcation surface.
This contribution is given by 
\be
V_1 = \int d\phi\int_{r_H}^{r_0} dr \ r \frac{l_{AdS}}{\sqrt{r^2 - r_H^2}}  = L \s{r_0^2 - r_H^2}.
\ee
The second one starts from the bifurcation surface and end on the ETW brane.
This is given by
\be
V_2= L\s{r_B^2 -r_H^2} = L r_H \f{Tl_{AdS}}{\s{1-(Tl_{AdS})^2}}  \label{eq:BehindHorizonV1}.
\ee
Therefore, the total holographic complexity for the volume is given by
\ba
C_V = \f{V_1 + V_2}{G_N l_{AdS}} &=& \f{L}{G_N l_{AdS}}\Big( \s{r_0^2 - r_H^2} +  r_H\f{Tl_{AdS}}{\s{1- (Tl_{AdS})^2}} \Big) \notag \\
&=& \f{2c}{3} \f{L}{\epsilon} \Big(\s{1 - \Big(\f{\pi \epsilon}{2 \delta}\Big)^2}+ \f{\pi \epsilon}{2 \delta} \f{Tl_{AdS}}{\s{1- (Tl_{AdS})^2}} \Big), \label{eq:cvtension}
\ea
where we translate the result using CFT quantities.
This is one of the main results in this paper.
It is interesting to note that (\ref{eq:BehindHorizonV1}) gives a temperature dependent contribution to complexity, though the contribution behind the horizon is finite for entanglement entropy.

In $T>0$ case, by taking $r_H \to r_0$ limit, we obtain
\be
C_V = \f{l_{AdS}}{G_N}\f{Tl_{AdS}}{\s{1-(Tl_{AdS})^2}}\f{L}{\ep}.
\ee
This is volume law divergent.

In $T<0$ case, when $r_H = \s{1 - (Tl_{AdS})^2}r_0$ the horizon comes in contact with the cutoff surface and entanglement entropy vanishes.
In this limit, the complexity becomes
\be
C_V = 0.
\ee 
Therefore, even in this case complexity becomes $0$ when entanglement entropy reduces to $0$.
 
\subsection{Action cases}
In the complexity equals action case, we consider the two ways of regularization of Wheeler de Witt patch, as noted in the section $3$.

\subsubsection{Regularization 1}
In this case, the configuration of the WdW patch depends on the horizon radius.
When $r_H > Tl_{AdS} r_0$, the null boundaries end on the black hole singularities.
This is always satisfied for negative tension case.
On the other hand, when $r_H < Tl_{AdS}r_0$ the null boundaries end on the ETW branes.
Here we consider both cases.

\begin{figure}[ht]
\begin{minipage}{0.5\hsize}
\begin{center}
\includegraphics[width=5cm]{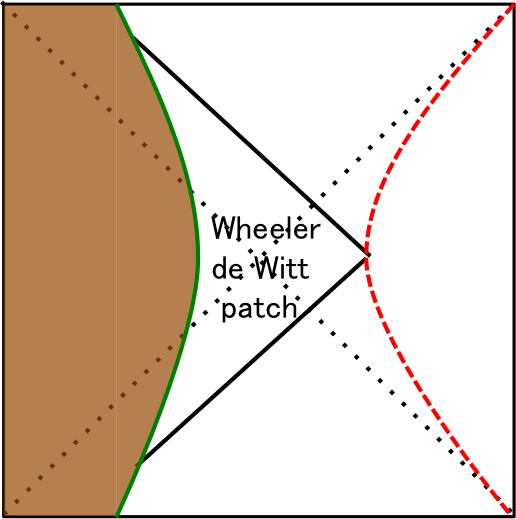}
\label{fig:rhlargec1}
\end{center}
\end{minipage}
\begin{minipage}{0.5\hsize}
\begin{center}
\includegraphics[width=5cm]{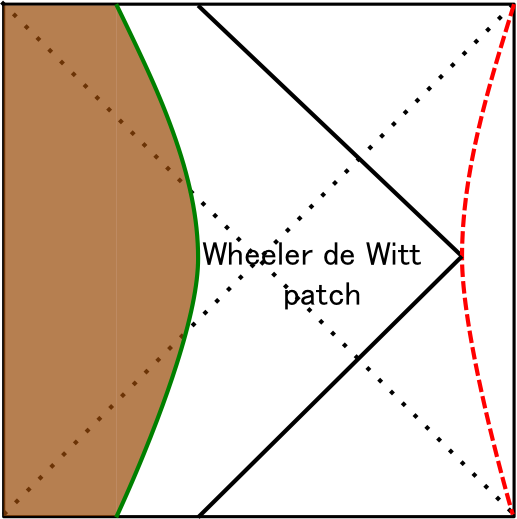}
\label{fig:rhsmallc2}
\end{center}
\end{minipage}
\caption{The left picture describes the configuration of WdW patch for $r_H>Tl_{AdS}r_0$.
The right picture is the one for $r_H<Tl_{AdS}r_0$.} 
\end{figure}

First we consider the case that  $r_H > Tl_{AdS}r_0$.
The Einstein-Hilbert term contribution becomes
\ba
S_{EH} &=& -\f{L}{4\pi G_N l_{AdS}} \Bigg[
r_H\f{(Tl_{AdS})^2}{1-(Tl_{AdS})^2} + (r_0 - r_H) \notag \\
&&+r_H\f{(r_0 - r_H)(1- 2(Tl_{AdS})^2) + 2  \s{r_0^2 - r_H^2 }Tl_{AdS} \s{1 - (Tl_{AdS})^2}}{r_0(1-(Tl_{AdS})^2)}
\Bigg].
\ea
The Gibbons-Hawking term contribution becomes
\ba
S_{GH} = \f{L}{4\pi G_N l_{AdS}} \Bigg[
r_H\f{(Tl_{AdS})^2}{1-(Tl_{AdS})^2} +
r_H\f{-(r_0 - r_H) (Tl_{AdS})^2 + \s{r_0^2 - r_H^2} Tl_{AdS} \s{1 - (Tl_{AdS})^2}}{r_0(1-(Tl_{AdS})^2)}
\Bigg].
\ea
The joint term contribution becomes
\ba
S_{joint} &=&  \f{L}{8 \pi G_N l_{AdS}} r_H \f{\s{1 - (Tl_{AdS})^2}r_H -Tl_{AdS} \s{r_0^2 - r_H^2} }{\s{1 -(Tl_{AdS})^2 }r_0} \log (1 - (Tl_{AdS})^2)\f{r_0^2}{r_H^2} \f{\alpha \tilde{\alpha}l_{AdS}^2}{r_0^2 - r_H^2} \notag \\
&& - \f{L}{8 \pi G_N l_{AdS}} r_0 \log \f{\alpha \tilde{\alpha}l_{AdS}^2}{r_0^2 - r_H^2}. 
\ea
The counter term contribution becomes
\ba
S_{ct} &=& 
 -\f{L}{8 \pi G_N l_{AdS}}\Bigg ( 2 r_H \f{\s{1 - (Tl_{AdS})^2}r_H -Tl_{AdS} \s{r_0^2 - r_H^2} }{\s{1 -(Tl_{AdS})^2 }r_0}   \notag \\
 &&+ r_H \f{\s{1 - (Tl_{AdS})^2}r_H -Tl_{AdS} \s{r_0^2 - r_H^2} }{\s{1 -(Tl_{AdS})^2 }r_0} \log \f{\alpha \tilde{\alpha}l_{ct}^2 (1 -(Tl_{AdS})^2) r_0^2}{r_H^2(\s{1 - (Tl_{AdS})^2}r_H -Tl_{AdS} \s{r_0^2 - r_H^2})^2}  \Bigg) \notag \\
&&+ \f{L}{8 \pi G_N l_{AdS}} \Big (2 r_0 + r_0 \log\f{\alpha\tilde{\alpha} l_{ct}^2}{r_0^2} \Big).
\ea 
Therefore, the complexity becomes
\ba
C_A &=& \f{S_{EH}+ S_{GH}+S_{joint}+S_{ct}}{\pi} \notag \\
&=&\f{L}{8 \pi^2 G_N l_{AdS}}\Big[
r_0 \log \f{(r_0^2 - r_H^2) l_{ct}^2}{r_0^2 l_{AdS}^2} \notag\\
&&- r_H \f{\s{1 - (Tl_{AdS})^2}r_H -Tl_{AdS} \s{r_0^2 - r_H^2} }{\s{1 -(Tl_{AdS})^2 }r_0} \log \f{(r_0^2-r_H^2)^2 l_{ct}^2}{(\s{1 - (Tl_{AdS})^2} r_H - Tl_{AdS}\s{r_0^2 - r_H^2})^2 l_{AdS}^2}
\Big]\notag \\
&& \qquad \qquad \qquad \qquad \qquad \qquad \qquad \qquad \qquad \qquad \qquad \qquad \qquad (\text{for}\ r_0 < Tl_{AdS} r_H ) .
\label{eq:cutoffsmall}
\ea
The Einstein-Hilbert term and the Gibbons-Hawking term contributions are cancelled with parts of the counter term contribution and finally joint term and counter term contributions remain finally.

Next we consider the $r_H < Tl_{AdS} r_0$ case.
In this case, the Einstein Hilbert term contribution is 
\ba
S_{EH} &=& - \f{L}{4\pi G_N l_{AdS}} \Big[ 
\Big(r_0 - \f{r_H^2}{r_0} \Big)
 + r_H \f{1 + 2 Tl_{AdS}}{1 +  Tl_{AdS}} 
 + r_H \f{(Tl_{AdS})^2}{1-(Tl_{AdS})^2}  \notag \\ 
 &&+ \f{1}{2}\log\f{r_0-r_H}{r_0+r_H}
 + r_H\text{arctanh} (Tl_{AdS})
\Big].
\ea
The Gibbons-Hawking term  contribution is 
\be
S_{GH } = \f{L r_H}{4 \pi G_N l_{AdS}} \Big[ \f{(Tl_{AdS})^2}{1-(Tl_{AdS})^2} + \f{Tl_{AdS}}{1 + Tl_{AdS}}  \Big]
\ee
The joint term contribution is 
\ba
S_{joint} = -\f{L}{8\pi G_N l_{AdS}} r_0 \log \f{\alpha \tilde{\alpha} l_{AdS}^2}{r_0^2 - r_H^2}. 
\ea
The counter term contribution is 
\be
S_{ct} = \f{L}{8\pi G_N l_{AdS}}\Big( 2r_0 +  r_0 \log \f{\alpha \tilde{\alpha} l_{ct}^2}{r_0^2} \Big).
\ee  
Therefore, the complexity becomes
\ba
C_A &=& \f{S_{EH}+ S_{GH}+S_{joint}+S_{ct}}{\pi} \notag \\
&=& \f{L}{4\pi G_N l_{AdS}} \Big[ 
r_0 \log \f{r_0 l_{ct}}{\s{r_0^2 - r_H^2} l_{AdS}} -r_H \log \s{\f{r_0-r_H}{r_0+r_H}} - r_H - r_H\text{arctanh}(Tl_{AdS}) \Big] \notag \\
 && \qquad \qquad \qquad \qquad \qquad \qquad \qquad \qquad \qquad \qquad \qquad \qquad \qquad (r_0 > Tl_{AdS} r_H ). \label{eq:cutofflarge}
\ea

(\ref{eq:cutoffsmall}) and (\ref{eq:cutofflarge}) are one of main results in this paper.
Note that when $T= 0$, (\ref{eq:cutoffsmall}) reduces to the half of the results in eternal black holes (\ref{eq:EternalChoice1}).
Because the tension dependent term have a minus sign, they contribute negatively.
This is different behavior from the volume case (\ref{eq:cvtension}), in which the contribution from the tension dependent term is positive for positive tension $T>0$.
A relative sign of additional term compared to the volume can be observed in \cite{Fu:2018kcp}, in which the geometry behind the horizon at $t=0$ is given by the higher genus surface.

When we take the horizon radius $r_H$ to be the cutoff radius $r_0$, almost all contributions cancel and we obtain a simple expression:
\be
C_A = \f{L}{8 \pi G_N l_{AdS}}r_0 \log (1 - (Tl_{AdS})^2).
\ee 
This is the value of action behind the horizon.
Note that this expression does not depend on the choice of counter term scale $l_{ct}$.
This is because the counter term integral on horizon vanishes.
This also means that even we omit the counter term, the result does not depend on the choice of the affine parametrization on each null boundary.

When $r_H = r_0 \s{1-(Tl_{AdS})^2}$, the cutoff surface $r = r_0$ contacts with the ETW brane in Euclidean signature and in Lorentzian signature for negative tension ETW branes.
In this case, complexity reduces to 
\ba
C_A &= &\f{L}{8\pi G_N l_{AdS}} \Big[ 
r_0 \log \f{(Tl_{AdS})^2 l_{ct}^2}{ l_{AdS}^2} \notag \\
&&- r_0 [1 - (Tl_{AdS})^2(1 + \text{sgn}(T))] \log \f{(Tl_{AdS})^2 l_{ct}^2}{[1 - (Tl_{AdS})^2(1 + \text{sgn}(T))]^2 l_{AdS}^2} \Big].
\ea
where $\text{sgn}(T)$ is the sign of the ETW brane tension $T$. 
Especially for $T<0$ this vanishes.
This is because there are no spacetime within the Wheeler de Witt patch.
Both of entanglement entropy (\ref{eq:bdyentropy}) and the complexity reduce to $0$ in this limit.

\subsubsection{Regularization 2}
In this case, the result depend on the sign of tension of ETW brane.
When $T>0$, the null boundaries end on the black holes singularities.
On the other hand, $T<0$ case the null boundaries end on the ETW branes.

\begin{figure}[ht]
\begin{minipage}{0.5\hsize}
\begin{center}
\includegraphics[width=5cm]{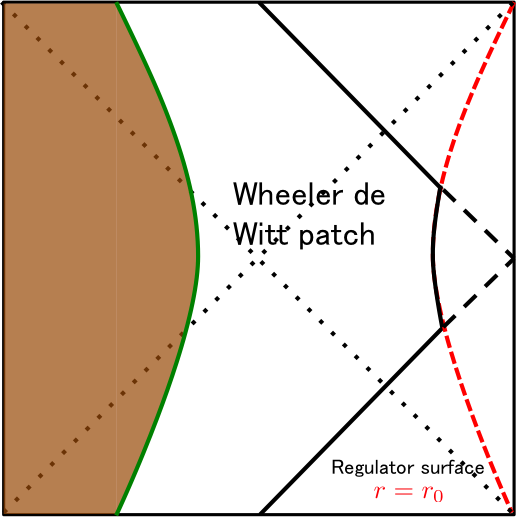}
\label{fig:rhlargec1}
\end{center}
\end{minipage}
\begin{minipage}{0.5\hsize}
\begin{center}
\includegraphics[width=5cm]{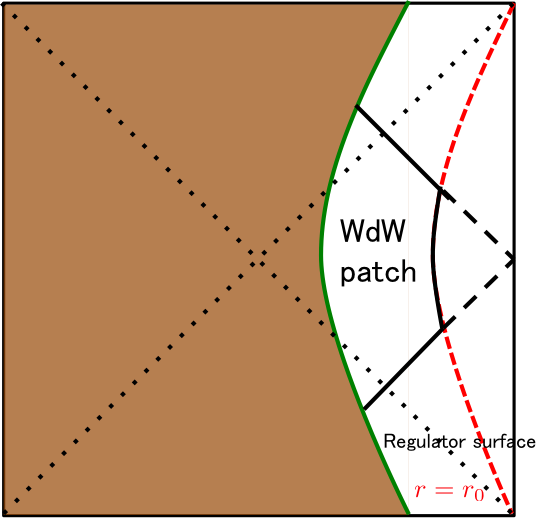}
\label{fig:rhsmallc2}
\end{center}
\end{minipage}
\caption{The left picture describes the configuration of the WdW patch for $T>0$.
The right picture is the configuration of the WdW patch for $T<0$.
$T=0$ cases can be considered as special limits of both cases.}
\end{figure}

First we consider the $T>0$ case.
the Einstein-Hilbert term contribution is 
\ba
S_{EH} &=& - \f{L}{4 \pi G_N l_{AdS}} \Big[ 
r_H \f{(Tl_{AdS})^2}{1- (Tl_{AdS})^2} + r_H \f{1 + 2 Tl_{AdS}}{1 + Tl_{AdS}} + r_H \text{arctanh}(Tl_{AdS}) \notag \\
&& + \f{2r_H\s{r_0 - r_H}}{\s{r_0 + r_H} + \s{r_0 - r_H}} - \f{\s{r_0^2 - r_H^2}}{\s{r_0 + r_H} - \s{r_0 - r_H}} \log \f{r_0 - r_H}{r_0 + r_H}.
\Big]
\ea
The Gibbons-Hawking term contribution is 
\ba
S_{GH} = \f{L}{4\pi G_N l_{AdS}} \Big[ r_H \f{(Tl_{AdS})^2}{1- (Tl_{AdS})^2} + r_H \f{ Tl_{AdS}}{1 + Tl_{AdS}} +  \f{2r_0^2 - r_H^2}{r_H} \log \s{\f{r_0 - r_H}{r_0 + r_H}} \Big].
\ea
The joint term contribution is 
\ba
S_{joint} = -\f{L}{4\pi G_N l_{AdS}}r_0 \log \f{\s{\alpha \tilde{\alpha } l_{AdS}}}{\s{r_0^2 - r_H^2}}.
\ea
The counter term contribution is 
\ba
S_{ct} = \f{L}{4\pi G_N l_{AdS}} \Big( r_ 0 + r_0 \log \f{\s{\alpha \tilde{\alpha } l_{ct}}}{r_0 }\Big).
\ea
Therefore, complexity becomes
\ba
C_A &=& \f{S_{EH} + S_{GH} + S_{joint} + S_{ct}}{\pi} \notag \\
&=&  \f{L}{4 \pi^2 G_N l_{AdS}}  \Big( (r_0-r_H) - \f{2r_H \s{r_0-r_H}}{\s{r_0+r_H} + \s{r_0-r_H}} + \f{r_H\s{r_0^2 -r_H^2}}{(\s{r_0+r_H} - \s{r_0-r_H})^2} \log \f{r_0-r_H}{r_0+r_H}  \notag \\ 
&&+ \f{(2r_0^2 - r_H^2)}{r_H} \log \s{ \f{r_0 + r_H}{r_0 - r_H} }  + r_0 \log\f{ \s{(r_0^2 -r _H^2)}l_{ct}}{r_0 l_{AdS}} - r_H \text{arctanh}(Tl_{AdS}) \Big).
\ea
When we take $T\to 0$, this reduces to the result of tensionless case (\ref{eq:t0a2}).
In $r_H \to r_0$ limit, this reduces to 
\be
C_A = \f{L}{4\pi^2 G_N l_{AdS}}r_0 \Big( \log\f{2l_{ct}}{l_{AdS}} -\text{arctanh}(Tl_{AdS}) \Big).
\ee
Note that this is proportional to $r_0$ and UV divergent.
The tension dependent term contributes negatively, and $\text{arctanh}(Tl_{AdS})$ can be arbitrarily large.

If $T< 0$, the null boundaries end on the ETW brane.
Then, the Einstein-Hilbert term contribution is 
\ba
S_{EH} &=& - \f{L}{4 \pi G_N l_{AdS}} \Bigg[
r_H \f{(Tl_{AdS})^2}{1-(Tl_{AdS})^2} + r_H \f{1 - 2(Tl_{AdS})^2 + 2 Tl_{AdS} \s{1-(Tl_{AdS})^2}}{1-(Tl_{AdS})^2} \notag \\
&&+\f{2\s{r_0 - r_H}}{\s{r_0 + r_H} + \s{r_0 - r_H}} - \f{r_H\s{r_0^2 - r_H^2}}{(\s{r_0+r_H}- \s{r_0-r_H})^2} \log \f{r_0 - r_H}{r_0+r_H}
\Bigg].
\ea
The Gibbons-Hawking term contribution is 

\ba
S_{GH} &=& \f{L}{4 \pi G_N l_{AdS}} \Big[ 
r_H \f{(Tl_{AdS})^2}{1-(Tl_{AdS})^2} + r_H \f{- (Tl_{AdS})^2 +  Tl_{AdS} \s{1-(Tl_{AdS})^2}}{1-(Tl_{AdS})^2} \notag \\
&&+ \f{2r_0^2 -r_H^2}{r_H} \log \s{\f{r_0 + r_H}{r_0-r_H}}  
\Big].
\ea
The joint term contribution is 
\ba
S_{joint} = - \f{L}{8 \pi G_N l_{AdS}} r_H \f{Tl_{AdS}}{\s{1 - (Tl_{AdS})^2}} \log \f{(1-  (Tl_{AdS})^2) \alpha \alpha 'l_{AdS}^2}{r_H^2} - \f{L}{8 \pi G_N l_{AdS}} r_0 \log \f{ \alpha \alpha 'l_{AdS}^2}{r_0^2 - r_H^2} .
\ea
The counter term term contribution is 
\ba
S_{ct} &=& \f{L}{8 \pi G_N l_{AdS}} \Big( 2r_H \f{Tl_{AdS}}{\s{1 - (Tl_{AdS})^2}} + r_H \f{Tl_{AdS}}{\s{1 - (Tl_{AdS})^2}} \log \f{(1-  (Tl_{AdS})^2) \alpha \alpha 'l_{ct}^2}{r_H^2 (Tl_{AdS})^2} \Big) \notag \\
&&+ \f{L}{8 \pi G_N l_{AdS}} \Big(2r_0 + r_0 \log \f{ \alpha \alpha ' l_{ct}^2}{r_0^2} \Big)
\ea
Therefore, the complexity is given by
\ba
C_A
 &=& \f{S_{EH} + S_{GH} + S_{joint} + S_{ct}}{\pi} \notag \\
&=& \f{L}{8\pi G_N l_{AdS}} \Big[ 2(r_0-r_H)
- \f{4 r_H\s{r_0 - r_H}}{\s{r_0+r_H} + \s{r_0-r_H}} 
\notag \\
&&+\f{2r_H\s{r_0^2 -r_H^2}}{(\s{r_0+r_H} - \s{r_0-r_H})^2} \log \f{r_0-r_H}{r_0+r_H}
+ \f{2r_0^2 -r_H^2}{r_H} \log \f{r_0 + r_H}{r_0 -r_H} \notag \\
 &&+ r_H \f{Tl_{AdS}}{\s{1-(Tl_{AdS})^2}} \log \f{(Tl_{AdS})^2 l_{ct}^2}{ l_{AdS}^2} + r_0\log \f{(r_0^2 - r_H^2)l_{ct}^2}{r_0^2l_{AdS}^2}
\Big].
\ea
When $r_H = r_0 \s{1 - (Tl_{AdS})^2}$, the cutoff surface comes in contact with the ETW brane  at $t=0$.
In this limit, the complexity becomes
\ba
C_A &=&\f{Lr_0}{8\pi G_N l_{AdS}} \Big[ 2  (1 - \s{1 - (Tl_{AdS})^2}) -\f{4 \s{1 - (Tl_{AdS})^2} (1 - \s{1 - (Tl_{AdS})^2})}{1 - T  -\s{1 - (Tl_{AdS})^2}} \notag \\
 && -\f{Tl_{AdS}\s{1- (Tl_{AdS})^2}}{1 + Tl_{AdS}} \log \f{1 - \s{1 - (Tl_{AdS})^2}}{1 + \s{1 - (Tl_{AdS})^2}} 
+\f{1 + (Tl_{AdS})^2}{\s{1 - (Tl_{AdS})^2}} \log \f{1 + \s{1 - (Tl_{AdS})^2}}{1 - \s{1 - (Tl_{AdS})^2}} \notag \\
&& + (1 +Tl_{AdS} ) \log \f{(Tl_{AdS})^2 l_{ct}^2}{l_{AdS}^2}.
\Big]
\ea
In this regularization, even in $r_H = r_0\s{1- (Tl_{AdS})^2}$ limit the complexity does not reduces to $0$ though entanglement entropy (\ref{eq:bdyentropy}) reduces to $0$.
This is because in this regularization the WdW patch does not vanish as is the case with the $T=0$ result(\ref{eq:t0a2}).
We found that in $T\to 0$ limit reduces to the $T=0$ results (\ref{eq:t0a2limit}).

\section{Conclusion and Discussion}
In this paper we studied the volume and the action, which are conjectured to be duals of the complexity in CFTs, both in eternal black holes and pure state black holes in AdS$_3$/BCFT$_2$ setup proposed in \cite{Takayanagi:2011zk}\cite{Fujita:2011fp}.
In this setup, the geometry havs end of the world (ETW) branes which have tension $T$.
We studied the tension dependence of the volume and the WdW action for $t=0$ carefully.
When the horizon radius $r_H$ and cutoff radius $r_0$ satisfy $r_H = r_0\s{1 - (Tl_{AdS})^2}$, the ETW brane comes in contact with the cutoff surface for negative tension in Lorentzian signature and for both sign of tension in Euclidean signature. 
If tension is positive, the horizon contacts to the cutoff surface when $r_H = r_0$.
We studied the behavior of the volume and the action in these limits.

We found that the volume increases when tension is positive $(T>0)$ and decreases when tension is negative $(T<0)$.
When the ETW brane comes in contact with the cutoff surface, volume reduces to $0$.
Because entanglement entropy also reduces to $0$ in this limit, a CFT dual of the volume becomes $0$ on product states.
Through the complexity = volume conjecture\cite{Susskind:2014rva}\cite{Stanford:2014jda}, this suggests that a reference state of the complexity is a product state.

The Wheeler de Witt patch can be regularized in different ways.
We studied two regularizations that are proposed in \cite{Carmi:2016wjl}.
For both cases, we give the analytic form of total action at $t=0$.
In the first regularization, in which the WdW patch ends on the cutoff surface, the total action vanishes when the horizon or the ETW brane contact with the cutoff surface.
Through the complexity=action conjecture\cite{Brown:2015bva}\cite{Brown:2015lvg}, this suggests that the reference state of complexity is a product state because entanglement entropy also reduces to $0$, which means the state is a product state.
This $0$ complexity are achieved in the geometry with $T=0$ ETW branes, which have a string/M theory realization\cite{Horava:1996ma}.
In the second regularization, in which the WdW patch ends on the asymptotic AdS boundary and then cutoff surface are introduced, the total action does not vanish even when the ETW brane or the black hole horizon contact with the cutoff surface but have UV divergence.
This is because the WdW patch does not vanish in this limit.
In this regularization, the reference state are taken to be a different state from the states that we studied.
It is an interesting future work to study which states leads to $0$ complexity in this regularization.
We expect that our method to study a product state limit in gravity side will be useful to study other geometric quantities which are duals of information theoretic quantities in CFTs.


There are several future problems.
In this paper,  we only consider the volume and the action at $t=0$.
The time dependence of the volume and the action with the regularization in which the WdW patch end on the asymptotic AdS boundary, this is done in \cite{Cooper:2018cmb}.
The time dependence with finite cutoff is also studied in \cite{Akhavan:2018wla}.
Because we only see the leading of $1/N$ expansion, it is interesting to study the subleading term, which is quantum correction in gravity, for both in holographic entanglement entropy and holographic complexity. 
It is interesting to study the time dependence of the volume and action for boundary states in the limit that the ETW brane or the black horizon contact with the cutoff surface.
Another interesting problem is the complexity for boundary states in Nearly AdS$_2/$Nearly CFT$_1$ setup.
In \cite{Kourkoulou:2017zaj}, boundary states for the SYK model are proposed.
2$d$ dilaton gravity solutions which share similar properties with SYK boundary states are also considered.
It is interesting future problem to study the complexity in this setup.

\subsection*{Note added:}
When this paper was in the final stage, the paper \cite{Cooper:2018cmb} appeared on the arXiv, in which they also computed the volume and the action in the AdS/BCFT setup.
The paper \cite{Akhavan:2018wla} also appeared on the arXiv, in which they argue the effect of cutoff on the time evolution of the complexity.

\section*{Acknowledgements}
We would like to thank
Alex Maloney, Satoshi Yamaguchi, Tadashi Takayanagi, Rob Myers, Henry Maxfield and Jamies Sully for helpful discussions. 
TN is supported by JSPS fellowships and the Simons Foundation through the It From Qubit collaboration.

\appendix
\section{Boundary states in $1+1$ dimensional Conformal Field Theory}
We summarize the basic properties of $1+1$ dimensional Boundary Conformal Field Theory (BCFT).
In BCFT, we put CFT on a manifold with boundaries.
We impose a perfect reflection condition to the energy momentum tensor on boundaries, which means that we keep the half of the conformal symmetry.
\subsection{notation of conformal field theory}
In this subsection we summarize the notation of 2d conformal field theories.
In 2d CFTs, the conformal symmetry $SO(2,2) \simeq SL(2,\mathbb{R})_L \times SL(2,\mathbb{R})_R$ is enhanced to an infinite dimensional symmetry called as Virasoro symmetry.
The generators of them are denoted as $\{L_n\}_{n \in \mathbb{Z}}$ for left-moving sector  and $\{\tilde{L}_n\}_{n \in \mathbb{Z}}$ fpr right-moving sector.
The commutation relation of them are given by 
\begin{align}
[L_n, L_m] = (n-m) L_{n+m} + \frac{c}{12}(n^3 - n) \delta_{n+m,0}, \notag   \\
[\tilde{L}_n, \tilde{L}_m] = (n-m) \tilde{L}_{n+m} + \frac{c}{12}(n^3 - n) \delta_{n+m,0} ,
\end{align}
where $c$ is the central charge of a given 2d CFT.
These generators are related to the energy momentum tensors.
The energy momentum tensor is expanded as 
\be
T(z) = \sum_{n\in \mathbb{Z}} L_n z^{-n-2}, \qquad \tilde{T}(\bar{z}) = \sum_{n\in \mathbb{Z}} \tilde{L}_n \bar{z}^{-n-2}
\ee
The vacuum state is given by 
\be
L_n\ket{0}  = \tilde{L}_n \ket{0} = 0 ,\qquad (n \ge -1).
\ee
and this commutation relations are expressed as the OPE of the energy momentum tensor:
\ba
T(z)T(w) \sim \frac{c}{2(z-w)^4} + \frac{2  T(w)}{(z-w)^2} + \frac{ \partial T(w)}{z-w} + \cdots \notag \\
\tilde{T}(\bar{z})\tilde{T}(\bar{w}) \sim \frac{c}{2(\bar{z}-\bar{w})^4} + \frac{2  \tilde{T}(\bar{w})}{(\bar{z}-\bar{w})^2} + \frac{ \partial \tilde{T}(\bar{w})}{\bar{z}-\bar{w}} + \cdots
\ea
A highest weight state of the Virasoro algebra is denoted as $\ket{h}_L$ for left-moving sector and $\ket{\bar{h}}_R$ for right-moving sector. 
The vector $\ket{h,\bar{h}} = \ket{h}_L \otimes \ket{\bar{h}}_R $ in the full Hilbert space is called as a primary state and that satisfies
\ba
&&L_n \ket{h,\bar{h}} = \tilde{L}_n \ket{h,\bar{h}} = 0 ,\qquad  (n > 0) \notag \\
&&L_0\ket{h,\bar{h}} = h \ket{h,\bar{h}}, \qquad \tilde{L}_0\ket{h,\bar{h}} = \bar{h}\ket{h,\bar{h}}.
\ea
The descendant states are constructed as 
\be
\cdots (L_{-n})^{k_n} (L_{-(n-1)})^{k_{n-1}} \cdots (L_{-1})^{k_1} \ket{h}_L 
\ee
and similarly for right-moving sector.
Then, we can choose an orthonormal basis $\ket{\vec{k},h}_L$, where $\vec{k} = (k_1,k_2, \cdots)$ is an infinite dimensional vector and all components satisfy $k _i \in \mathbb{Z}_+$.
The Hilbert space spanned by these vectors is denoted as $\mathcal{H}_h$ for left-moving modes and similarly 
$\bar{\mathcal{H}}_{\bar{h}}$.
The full Hilbert space is given by $\mathcal{H} =\sum_{h,\bar{h}}\mathcal{M}_{h,\bar{h}}\mathcal{H}_h\otimes \bar{\mathcal{H}}_{\bar{h}}$.
The matrix $\mathcal{M}_{h,\bar{h}}$ is chosen to satisfy the modular invariance on a torus.
When $\mathcal{M}_{h,\bar{h}} = \delta _{h,\bar{h}}$,  this model is called as a diagonal model.
In this paper we focus on these cases.

\subsection{construction of boundary states}
Let us consider a semi infinite cylinder which is a simple example of manifolds with boundaries.
For simplicity, we only consider diagonal modular invariant models.
This cylinder can be conformally mapped to a disk.
In the open string picture, we think that this boundary is located on a space and the boundary is static (in  Euclidean signature).
The Hilbert space is defined on a half line with the boundary in this picture.
In closed string picture, we think that this boundary suddenly appears in (Euclidean) time.
The Hilbert space is defined on a circle in this picture, so there are no change in the Hilbert space.
The existence of boundary is expressed as a state in the closed string picture.
This is so called a boundary state.

We keep the half of the conformal symmetry.
This condition is expressed as 
\be
(T(z) - \tilde{T}(\bar{z}))|_{|z| = 1} \ket{B} = 0.
\ee
By applying the contour integral $\oint _{|z| = 1}$, we obtain
\be
(L_n - \tilde{L}_{-n})\ket{B} = 0. \label{eq:condition1}
\ee
In each sector, we can find a simple algebraic solution of (\ref{eq:condition1}):
\be
| I_h \rangle\rangle = \sum_{\vec{k}} \ket{\vec{k},h}_L \otimes \ket{\vec{k},h}_R.
\ee
This is called as the Ishibashi state for the primary state $\ket{h,h}$\cite{Ishibashi:1988kg,Onogi:1988qk}.
Ishibashi states are maximally entangled state in each conformal sector.
Actual boundary states that corresponds to physical boundary conditions are so called Cardy states\cite{CARDY1989581}.
Cardy states are given by the sum of Ishibashi states:
\be
\ket{B_{\alpha}} = \sum_{h} \mathcal{B}_{\alpha}{}^h| I_h \rangle\rangle
\ee
The Cardy states satisfy the consistency condition for partition functions on finite cylinders,
which means that the same answer is obtained from the open string picture and the closed string picture.
Open string partition function is given by $Z_{\alpha,\beta}^{open}(t) = \sum_{h} n_{\alpha,\beta}^h \chi_h (e^{-t})$ where $n_{\alpha,\beta}^h$ is a non negative integer and $\chi_h(q)$ is a character for the sector $h$.
Open-closed duality says that this partition function should be related to the amplitude in boundary states via the modular transformation.
This condition is written as 
\be
\bra{B_{\alpha}} e^{- l \cdot H} \ket{B_{\beta}} = \sum_{h,h'} n_{\alpha,\beta}^h \mathcal{S}_h{}^{h'} \chi_{h'}(e^{-t}),
\ee
where $\mathcal{S}_h{}^{h'}$ is the modular $S$ matrix and $tl = \pi$.
This gives constraint on the coefficient $\mathcal{B}_{\alpha}{}^h$ and the solutions give the physical boundary conditions.
Solving these constraints is a hard problem.
In rational CFTs, we can derive the solutions using the modular S matrices.
We can also solve this modular bootstrap problem in Liouville field theory.
The AdS/BCFT setup, on the other hand, satisfies two conditions and they give physical (Cardy) boundary states holographically.

\section{Coordinates in AdS$_3$}
In this appendix we summarize the coordinates of AdS$_3$ and BTZ black holes.
The embedding space of AdS$_3$ is given by
\be
-T_1^2 -T_2^2 + X_1^2 + X_2^2 = -l_{AdS}^2,
\ee
and metric is induced from 
\be
ds ^2 = -dT_1^2 - dT_2^2 +dX_1^2 +dX_2^2. \label{eq:embed}
\ee
A coordinate that covers the global AdS$_3$ is 
\ba
T_1&=& l_{AdS} \cosh \chi \cos t_g  ,\qquad
T_2=l_{AdS} \cosh \chi \sin t_g \notag  \\
X_1&=&l_{AdS} \sinh \chi \sin\varphi,\qquad
X_2=l_{AdS} \sinh \chi \cos \varphi ,
\ea
The metric for this global coordinate is given by
\be
ds^2  = l_{AdS}^2(- \cosh^2 \chi dt_g^2+ d\chi^2 + \sinh^2 \chi d \varphi^2).
\ee
By changing the coordinate $\cosh\chi = 1/\cos\theta ( \sinh \chi = \tan \theta)$, we obtain the metric that is manifestly conformally equivalent to the Einstein static universe:
\be
ds^2 = \f{l_{AdS}^2}{\cos^2\theta} (- dt_g^2 + d \theta^2 + \sin^2 \theta d \varphi^2)
\ee
The Poincare coordinate of AdS$_3$ is given by 
\ba
T_1&=& \f{1}{2z}(l_{AdS}^2+(z^2 -x_0^2 +x_1 ^2) )  ,\qquad
T_2=l_{AdS} \f{x_0}{z}  \notag  \\
X_1&=&l_{AdS} \f{x_1}{z},\qquad
X_2=\f{1}{2z}(l_{AdS}^2-(z^2 -x_0^2 +x_1 ^2) )  , 
\ea
and metric becomes 
\be
ds ^2 = R^2\  \f{dz^2 -dx_0^2 +dx_1^2 }{z^2}.
\ee
This coordinate covers only the half of the global AdS$_3$ defined by $T_1 + X_1 > 0$.

To obtain the metric of BTZ black holes, we use the following embedding:
\ba
T_1&=& l_{AdS} \f{v+u}{1+uv} = \f{l_{AdS}}{r_H} \s{r^2 - r_H^2} \sinh\f{r_H t}{l_{AdS}^2} =  \f{l_{AdS}}{r_H} \s{r_H^2-r^2 } \cosh\f{r_H \tilde{t}}{l_{AdS}^2} ,\notag \\
T_2&=&l_{AdS} \f{1-uv}{1+uv} \cosh\f{r_H \phi}{l_{AdS}} = \f{r}{r_H}\cosh\f{r_H \phi}{l_{AdS}}  \notag  \\
X_1&=& l_{AdS}\f{v-u}{1+uv} =\f{l_{AdS}}{r_H} \s{r^2 -r_H^2} \cosh\f{r_H t}{l_{AdS}^2}  = \f{l_{AdS}}{r_H} \s{r_H^2-r^2 } \sinh\f{r_H \tilde{t}}{l_{AdS}^2} ,\notag \\
X_2&=& l_{AdS}\f{1-uv}{1+uv} \sinh\f{r_H \phi}{l_{AdS}} = \f{r}{r_H}\sinh\f{r_H \phi}{l_{AdS}} \label{eq:BTZembed} 
\ea
Then, the metric for the eternal BTZ black hole is given by
\be
ds^2 = -\frac{4l_{AdS}^2 }{(1 + uv)^2} du dv + r_H^2 \Big( \f{1-uv}{1+uv} \Big)^2 d\phi^2.
\ee
with $ -1 < u,v < 1$.

$(r,t,\phi)$ coordinate only covers the right exterior of BTZ black holes.
The metric for outside of the BTZ black hole is given by 
\be
ds^2 = - \f{r^2 - r_H^2}{l_{AdS}^2} dt ^2 + \frac{l_{AdS}^2}{r^2 - r_H^2} dr^2 + r^2 d \phi^2.
\ee
By analytic continuation of time $t$, we can go to the inside or the left exterior.
For example, we can go to the future interior by continuing $t = \tilde{t} - i\beta/4$ for real $\tilde{t}$ and the Hawking temperature $T_H = 1/\beta$. 
The metrics for interiors take the same form with that for exteriors, but the roles of time and radial direction are exchanged. 

\section{Brief summary of AdS/BCFT}
We briefly summarize AdS/BCFT proposed in \cite{Takayanagi:2011zk}\cite{Fujita:2011fp}  .
The general idea is to consider another boundary that is the extension of boundary in CFT to the AdS bulk.
This second boundary can be seen as the end-of-the world (ETW) brane that emanate from the asymptotic boundary.
We call AdS bulk as $N$, asymptotic AdS boundary $M$, and ETW brane $Q$.
The end point of ETW brane $Q$ on AdS boundary $M$ is exactly the same with the location of boundary in CFT.
The general action for this holographic setup is
\be
S =  \f{1}{16\pi G_N} \int_N  \s{-g}(R - 2 \Lambda) + \f{1}{8\pi G_N}\int _M\s{-h} K + \f{1}{8\pi G_N}\int _Q\s{-h} K + \int _Q \s{-h} \mathcal{L}_{matter}.
\ee
The last two terms are new terms in this setup.

As usual, we impose Dirichlet boundary condition on the asymptotic boundary $M$.
On the other hand, we impose Neumann boundary condition on the ETW brane $Q$.
This can be seen as a natural extension of a system with gravity and point particles which are dynamical and backreact to gravity.
Here the ETW brane is dynamical and can backreacts to gravity.
Now, the well defined variational principle impose the following boundary equation of motion:
\be
K_{ab} - K h_{ab} = 8 \pi G_N T^Q_{ab}, \label{eq:bdyeom}
\ee
where we defined the boundary matter stress energy tensor $T^{Qab} = \f{2}{\s{-h}}\f{\delta I_Q}{\delta h_{ab}}$ for $I_Q = \int _Q \s{-h} \mathcal{L}_{matter}$.

Now, we move to a simple example with matter Lagrangian $\mathcal{L}_{matter} = -\f{T}{8\pi G_N}$.
In other word, we consider the situation that the ETW brane has only tension $T$.
Now, the boundary equation of motion (\ref{eq:bdyeom}) becomes
\be
K_{ab} =(K-T)h_{ab}. \label{eq:bdyeombrane}
\ee
The trace of this gives $K = \f{d}{d-1}T$.

In poincare coordinate, we can give a simple solution of this equation.
To realize the symmetry of BCFT, we put thef following ansatz\footnote{By using conformal symmetry of BCFT, we can determine the metric up to the following form 
$ds^2 = d\rho^2 + e^{2A(\rho)} ds_{AdS_d}^2$,
where $A(\rho)$ is a function of $\rho$.
In other words, the warp factor which depend on $A(\rho)$ is not determined by the isometry $SO(d-1,2)$ that corresponds to the conformal symmetry of BCFT.}:
\be
ds^2 = d\rho^2 + \cosh^2\f{\rho}{l_{AdS}} ds_{AdS_d}^2, \label{eq:AdSslice}
\ee
If we assume that $-\infty < \rho < \infty$, then the above metric is equivalent to the $AdS_{d+1}$ metric.
This can be confirmed by putting $ds_{AdS_d}^2 = l_{AdS}^2 \f{-dx_0 ^2 + dy^2 + d\vec{w}^2}{y^2}, (\vec{w}   \in \mathbb{R}^{d-2})$  and doing the coordinate transformation from $(\rho,y)$ to $(z,x_1)$ which is given by
\be
z =  \frac{y}{\cosh\f{\rho}{l_{AdS}}} , \qquad x_1 = y \tanh \f{\rho}{l_{AdS}}, \label{eq:maptousual}
\ee
we obtain the $AdS_{d+1}$ metric in Poincare coordinate:
\be
ds^2 = l_{AdS}^2 \f{-dx_0^2 + dz^2 + dx_1^2 + d\vec{w}^2}{z^2}. \label{eq:usualPoincare}
\ee
Now, we consider the $\rho = \rho_*$ slice.
It is easy to derive the extrinsic curvature of this slice\footnote{When the metric take the form $ds^2 = dr ^2 + h_{ab}dx^adx^b$, then the extrinsic curvature on $r = \text{const}$ slice is given by $K_{ab} = \f{1}{2}\f{\partial h_{ab}}{\partial r}$.}:
\be
K_{ab} = \f{1}{l_{AdS}} \tanh\f{\rho}{l_{AdS}} h_{ab}.
\ee
Therefore the extrinsic curvature on this slice is proportional to the induced metric and can give a solution of (\ref{eq:bdyeombrane}).
By putting this to (\ref{eq:bdyeombrane}),  $\rho_*$ is determined in terms of the brane tension $T$:
\be
T = \f{d-1}{l_{AdS}} \tanh \f{\rho_*}{l_{AdS}}.
\ee

In Poincare patch, ETW branes are located at $\rho = \rho_*$ slice in (\ref{eq:AdSslice}) coordinate.
In the ordinary coordinate (\ref{eq:usualPoincare}), ETW branes are located at 
\be
\f{x_1}{z} = \sinh\f{\rho*}{l_{AdS}}
\ee

Now we concentrate on AdS$_3$ cases and construct pure boundary state black holes.
In three dimensional gravity, vacuum solutions are always locally AdS$_3$.
Actually, BTZ black holes are given by the orbifolds of global AdS$_3$.
The embedding is given by (\ref{eq:BTZembed}).

we obtain the trajectory of the ETW brane:
\be
\f{Tl_{AdS}}{\s{1-(Tl_{AdS})^2}} = \f{\s{r(t)^2-r_H^2}}{r_H} \cosh \f{r_Ht}{l_{AdS}^2}.
\ee
Induced metric on this coordinate is given by
\be
ds_{brane}^2 = - \f{r_H^4}{l_{AdS}^2}  \Big(\f{Tl_{AdS}}{1 - (Tl_{AdS})^2}\Big) ^2  \f{1}{r(t)^2}\f{1}{\cosh^4 \f{r_H t}{l_{AdS}^2}}  dt ^2 + r(t)^2 d\phi^2.
\ee
To get the Euclidean configuration, first we move to right exterior by $t\to -t + i\f{\beta}{2}$, by analytically continuing $ \tau = -i t$ we obtain
\be
\f{Tl_{AdS}}{\s{1-(Tl_{AdS})^2}} =  -\f{\s{r(\tau)^2-r_H^2}}{r_H} \cos \f{r_H\tau}{l_{AdS}^2}.
\ee
This is the configuration of ETW brane in Euclidean BTZ black holes\cite{Almheiri:2018ijj}.
Especially we find that this ETW brane ends on the antipodal points of the thermal circle.
These end points are identified with the sudden appearance of the boundary in Euclidean time in path integral preparation of regularized boundary states $e^{-\delta \cdot H} \ket{B}$.


In the Kruskal coordinate, the configuration of the ETW brane is given by
\be
X_1 = \f{v-u}{1+uv} = - \frac{Tl_{AdS}}{\s{1 - (Tl_{AdS})^2}}.
\ee
This can be written as
\be
u(v) = \frac{\sqrt{1 -(Tl_{AdS})^2 } v + Tl_{AdS}}{\sqrt{1 -(Tl_{AdS})^2 } - Tl_{AdS} v}. \label{eq:utrajectory}
\ee
The derivative becomes
\be
\f{du}{dv} = \f{1}{(\s{1 - (Tl_{AdS})^2} - Tl_{AdS}v)^2}.
\ee
The induced metric on the ETW brane is given by
\be
ds_{brane}^2 = - \frac{4 l_{AdS}^2  }{[1-(Tl_{AdS})^2](1 + v^2 )^2 } dv^2 + r_H^2 \frac{[\sqrt{1 -(Tl_{AdS})^2 } (1 - v^2 ) - 2Tl_{AdS} v ]^2}{[1-(Tl_{AdS})^2](1 + v^2 )^2} d\phi^2.
\ee

In the interior, the location of the ETW brane is given by
\be
\f{Tl_{AdS}}{\s{1-(Tl_{AdS})^2}} = \f{\s{r_H^2-r(t)^2}}{r_H} \sinh \f{r_H\tilde{t}}{l_{AdS}^2},
\ee
and the induced metric on the ETW brane becomes
\be
ds_{brane}^2 = - \f{r_H^4}{l_{AdS}^2}  \Big(\f{Tl_{AdS}}{1 - (Tl_{AdS})^2}\Big) ^2  \f{1}{r(\tilde{t})^2}\f{1}{\sinh^4 \f{r_H \tilde{t}}{l_{AdS}^2}}  dt ^2 + r(\tilde{t})^2 d\phi^2.
\ee

\section{detail of computation of the action}
\begin{figure}[ht]
\begin{minipage}{0.5\hsize}
\begin{center}
\includegraphics[width=6cm]{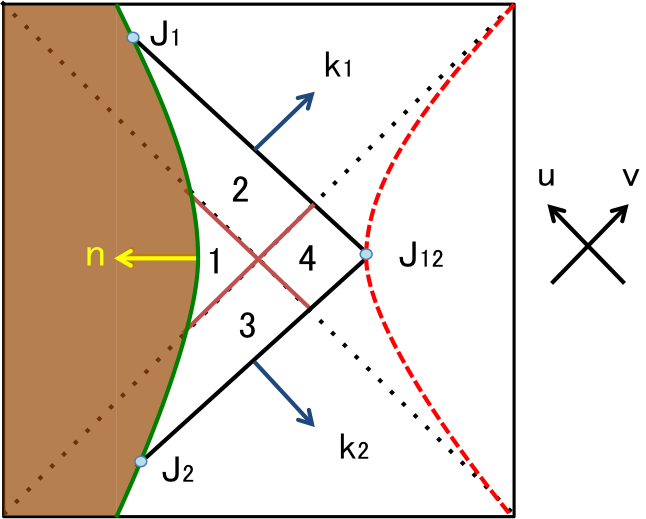}
\label{fig:actionapp1}
\end{center}
\end{minipage}
\begin{minipage}{0.5\hsize}
\begin{center}
\includegraphics[width=5cm]{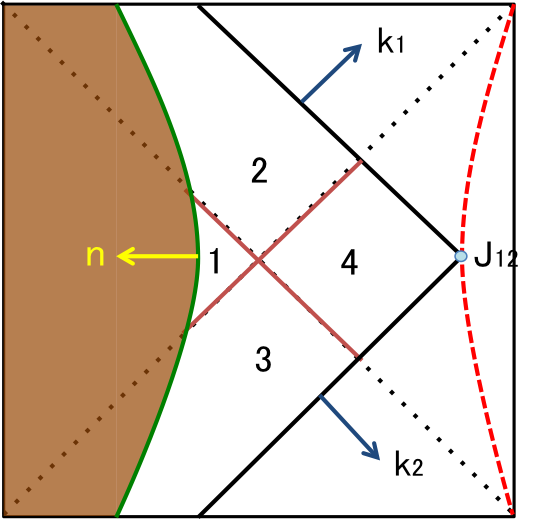}
\label{fig:actionapp2}
\end{center}
\end{minipage}
\caption{The left picture is the configuration for regularization 1 with positive tension ETW branes for $r_H > Tl_{AdS}r_0$.
The calculation for this configuration is also applicable for negative tension cases.
The right picture is the configuration for regularization 1 with positive tension ETW branes for  $r_H < Tl_{AdS}r_0$.  
If we decrease the temperature further, we will see the Hawking-Page transition. }
\label{fig:actioncalculation}
\end{figure}

In this section, we show some detail calculations of the action for the regularization $2$ case where the WdW patch ends on the cutoff surface.. 
It is convenient to divide the regions as in the Figure\ref{fig:actioncalculation}.

\subsection{Einstein Hilbert term and Gibbons Hawking term}
We first consider the interior contribution.
The contribution from the left region is 
\ba
\int_{region1} \s{-g} = \int d \phi \int_{- \infty}^{\infty} dt \int _{r_H}^{r(t)} dr \ r &=& \frac{1}{2} \frac{L}{l_{AdS}} \int_{- \infty}^{\infty} dt\ (r(t)^2 - r_H^2) \notag \\
&=& \frac{1}{2} \frac{L}{l_{AdS}} \frac{r_H^2 (Tl_{AdS})^2}{1 -  (Tl_{AdS})^2} \int_{- \infty}^{\infty} dt \frac{1}{\cosh^2 (r_H t/l_{AdS}^2)} \notag \\
&=&  r_H l_{AdS} L  \frac{ (Tl_{AdS})^2}{1 -  (Tl_{AdS})^2} 
\ea

The contribution from boundary 1 is
\ba
\int_{boundary1} \s{-\gamma} &=& \f{r_H^2}{l_{AdS}} \frac{Tl_{AdS}}{1 -  (Tl_{AdS})^2}   \int _{-\infty}^{\infty}dt   \frac{1}{\cosh^2\f{r_H t }{l_{AdS}^2}} \int d \phi \notag \\
&=& 2 r_H L \frac{ Tl_{AdS}}{1 -  (Tl_{AdS})^2}
\ea 
The contribution from the region 2 and 3 is
\ba
\int_{region2 } \s{-g} &=&2r_H l_{AdS}^2  \int _0 ^a dv \int _0 ^{u(v)} du \f{1 - uv }{(1 + uv)^3} \int d\phi \notag \\
&=& 2 r_H l_{AdS} L \int _0 ^a \f
{(\s{1 - (T l_{AdS})^2} - Tl_{AdS}v)(\s{1 -(T l_{AdS})^2 } v + T l_{AdS})}
{(1 -  (Tl_{AdS})^2) (1 + v^2)} \notag \\
&=& 2 r_H l_{AdS} L \f
{2 a Tl_{AdS}\s{1 - (T l_{AdS})^2} + a^2 (1 - 2 (Tl_{AdS})^2) }
{2(1+a^2)((1 -  (Tl_{AdS})^2)}
\ea
The contribution from the boundary 2 and 3 is
\ba
\int_{boundary 2} \s{-\gamma} &=& 2 r_H l_{AdS}\int_0^a dv \int d\phi \f
{\s{1 - (T l_{AdS})^2} (1 - v^2) - 2T l_{AdS} v }
{(1 -  (Tl_{AdS})^2) (1 + v^2)^2} \notag \\
&=& 2 r_H L \f
{a \s{1 - (T l_{AdS})^2} - a^2T l_{AdS} }
{(1 + a^2)(1 -  (Tl_{AdS})^2)}
\ea
The contribution from the region 4 is 
\ba
\int_{region4} \s{-g} &=& 2r_H l_{AdS}^2  \int _0 ^a dv \int _{-a} ^0 \f{1 - uv }{(1 + uv)^3} \int d\phi \notag \\
&=& 2 r_H l_{AdS} L \f{a^2}{1-a^2}
\ea

If $a> v_* =\s{\f{1-Tl_{AdS}}{1+Tl_{AdS}}}$, the contribution from the region2 and 3 is changed as
\ba
\int_{region2 } \s{-g} &=&2r_H l_{AdS}^2  \int _0 ^{v_*} dv \int _0 ^{u(v)} du \f{1 - uv }{(1 + uv)^3} \int d\phi + 2r_H l_{AdS}^2  \int _{v_*} ^{a} dv \int _0 ^{v^{-1}} du \f{1 - uv }{(1 + uv)^3} \int d\phi \notag \\
&=& 2 r_H l_{AdS} L \f{1+ 2 Tl_{AdS}}{4 + 4 Tl_{AdS}} + \f{ r_H l_{AdS} L}{2}  ( \log a + \text{arctanh}(Tl_{AdS}) )
\ea

and contribution from the boundary 2 is 
\ba
\int_{boundary 2} \s{-\gamma} &=& 2 r_H l_{AdS}\int_0^{v_*} dv \int d\phi \f
{\s{1 - (T l_{AdS})^2} (1 - v^2) - 2T l_{AdS} v }
{(1 -  (Tl_{AdS})^2) (1 + v^2)^2} \notag \\
&=& 2r_H L \f{1}{2 + 2Tl_{AdS} }
\ea

\subsection{Joint terms with Null surfaces}
Joint terms are given by
\be
S_{joint} = \f{1}{8 \pi G_N} \int _J  d^{d-1} \theta  \s{\gamma}  A
\ee
where $A$ is given by $A_{12} =\log|\f{1}{2}\bm{k}_1 \cdot\bm{k}_2|$, $A_1 = -\log|\bm{k}_1 \cdot \bm{n}|$ or $A_2 = -\log|\bm{k}_2\cdot\bm{n}|$ at $J_{12}$, $J_1$ or $J_2$.
The 1 forms that are orthogonal to each surface is 
\be
\bm{k}_1  = \alpha (dt + \f{dr}{f(r)}) = \alpha \f{l_{AdS}^2}{r_H}\f{1}{v} dv 
\ee

\be
\bm{k}_2  = \tilde{\alpha} (-dt + \f{dr}{f(r)}) = -\tilde{\alpha} \f{l_{AdS}^2}{r_H}\f{1}{u} du
\ee

\be
\bm{n} = \f{l_{AdS}}{1 + u(v)v} \f{1}{u'(v)} (du - u'(v)dv).
\ee
where $u(v)$ is the one given in \ref{eq:utrajectory}.
\be
A_1 = -\log|\bm{k}_1 \cdot \bm{n}|  = \log \Big(\f{\s{1-(Tl_{AdS})^2}}{2} (1+a^2) \f{l_{AdS}}{r_H} \f{\alpha l_{AdS}}{a} \Big)
\ee

\be
A_2 = -\log|\bm{k}_2 \cdot \bm{n}|  =  \log \Big(\f{\s{1-(Tl_{AdS})^2}}{2} (1+a^2) \f{l_{AdS}}{r_H} \f{\tilde{\alpha} l_{AdS}}{a} \Big)
\ee

\be
A_{12} =  \log \f{1}{2} |\bm{k}_1 \cdot \bm{k}_2| = \f{1}{2} \log \f{\alpha \tilde{\alpha}}{4 a^2} \Big(\f{l_{AdS}}{r_H} \Big)^2 (1-a^2)
\ee

\subsection{Counter term on null boundaries}
On $v = v_0$ slice, counter term integral becomes
\be
S_{ct} =  \f{1}{8 \pi G_N} \int d \lambda d\phi \s{\gamma} \Theta \log (l_{ct} \Theta),
\ee
with 
\be
\f{\partial}{ \partial \lambda}  = -\alpha\f{(1 + uv_0)^2}{2 r_H v_0} \f{\partial}{\partial u}
\ee
and 
\ba
\Theta = \f{\partial}{ \partial \lambda} \log \s{\gamma} &=& -\alpha\f{(1 + uv_0)^2}{2 r_H v_0} \f{\partial}{\partial u}\log\Big(r_H \f{1 - uv_0}{1+uv_0} \Big) \notag \\
&=& \f{\alpha}{ r_H} \f{1 + uv_0}{1-uv_0}.
\ea
The integral becomes
\ba
S_{ct} &=& -\f{L }{8 \pi G_Nl_{AdS}}\int _{u_1} ^{u_2}d u \f{2 r_H v_0}{(1+uv_0)^2} \log \Big( \f{\alpha}{r_H} \f{1 + uv_0}{1-uv_0} \Big) \notag \\
&=& -\f{L}{8 \pi G_Nl_{AdS}}\Big[r_H \Big( \f{1 - uv_0}{1+uv_0}\Big) +r_H \Big( \f{1 - uv_0}{1+uv_0}\Big) \log \Big( \f{\alpha l_{ct} }{ r_H} \f{1 + uv_0}{1-uv_0} \Big)\Big]_{u_1}^{u_2} \notag  \\
&=& -\f{L}{8 \pi G_Nl_{AdS}}\Big[ r(u,v_0) + r(u,v_0) \log \f{\alpha l_{ct}}{r(u,v_0)}\Big]_{u_1}^{u_2}.
\ea
Here we put $r(u,v_0) = r_H \f{1 - uv_0}{1+uv_0}$ in the third line and $u_1,u_2$ are $u$ coordinate of two end points. 
The second term contains the logarithm of normalization $\alpha$ and this cancels the affine parametrization dependence of joint terms.
From this expression, it is easy to see that the counter term contribution vanishes on the horizon because on the horizon $r(u,v_0) = r_H$ and total contribution vanishes. 

\bibliography{complexitybunken}

\end{document}